\documentclass[useAMS,usenatbib]{mn2e}
\usepackage{natbib}
\usepackage{color,graphicx} 
\usepackage{amsmath}        
\usepackage{amsfonts}       
\usepackage{amssymb}        
\usepackage{gensymb}        
\usepackage{framed}         
\usepackage{upgreek}
\usepackage{xspace}
\usepackage{wrapfig}
\usepackage{rotating}
\usepackage[space]{grffile}
\usepackage{latexsym}
\usepackage{textcomp}
\usepackage{longtable}
\usepackage{multirow,booktabs}
\usepackage{url}
\usepackage{hyperref}
\newif\iflatexml\latexmlfalse
\usepackage[utf8]{inputenc}
\usepackage[english]{babel}
\usepackage{xcolor,listings}
\usepackage{textcomp}
\usepackage[T1]{fontenc}
\usepackage{ae,aecompl}
\usepackage{lmodern}

\author[Keane et al.]{E.F.~Keane$^{1,2,3}$, E.D.~Barr$^{2,4}$,
  A. Jameson$^{2,3}$, V. Morello$^{2,3}$, M. Caleb$^{5,2,3}$,
  S. Bhandari$^{2,3}$,\newauthor E. Petroff$^{6,2,3,7}$,
  A. Possenti$^{8}$, M. Burgay$^{8}$, C. Tiburzi$^{4,9}$,
  M. Bailes$^{2,3}$, N. D. R. Bhat$^{10,3}$, \newauthor
  S. Burke-Spolaor$^{11}$, R.P. Eatough$^{4}$, C. Flynn$^{2}$,
  F. Jankowski$^{2,3}$, S. Johnston$^{7}$, \newauthor
  M. Kramer$^{4,12}$, L. Levin$^{12}$, C. Ng$^{13,4}$, W. van
  Straten$^{2}$ \& V. Venkatraman Krishnan$^{2,3}$
\\ $^{1}$ SKA Organisation, Jodrell
  Bank Observatory, SK11 9DL, UK. \\ $^{2}$ Centre for Astrophysics
  and Supercomputing, Swinburne University of Technology, Mail H30, PO
  Box 218, VIC 3122, Australia. \\ $^{3}$ ARC Centre of Excellence for
  All-sky Astrophysics (CAASTRO). \\ $^{4}$ Max-Planck-Institut f\"ur
  Radioastronomie, Auf dem H\"ugel 69, D-53121 Bonn, Germany. \\
  $^{5}$ Research School of Astronomy and Astrophysics, Australian
  National University, ACT 2611, Australia. \\ $^{6}$ ASTRON, the
  Netherlands Institute for Radio Astronomy, Postbus 2, NL-7990 AA
  Dwingeloo, the Netherlands.\\ $^{7}$ CSIRO Astronomy \& Space
  Science, Australia Telescope National Facility, P.O. Box 76, Epping,
  NSW 1710, Australia. \\ $^{8}$ INAF --- Osservatorio Astronomico di
  Cagliari, Via della Scienza 5, I-09047 Selargius (CA), Italy. \\
  $^{9}$ Fakult\"{a}t f\"{u}r Physik, Universität Bielefeld, Postfach
  100131, D-33501 Bielefeld, Germany. \\ $^{10}$ International Centre
  for Radio Astronomy Research, Curtin University, Bentley, WA 6102,
  Australia. \\ $^{11}$ National Radio Astronomy Observatory, Socorro,
  NM, USA. \\ $^{12}$ Jodrell Bank Centre for Astrophysics, School of
  Physics and Astronomy, The University of Manchester, Manchester M13
  9PL, UK. \\ $^{13}$ Department of Physics and Astronomy, University
  of British Columbia, 6224 Agricultural Road, Vancouver, BC V6T 1Z1,
  Canada \\} \date{\today} \title[SUPERB I]{The SUrvey for Pulsars and
  Extragalactic Radio Bursts I: Survey Description and Overview}
\begin{document}

\maketitle

\begin{abstract}

  We describe the Survey for Pulsars and Extragalactic Radio Bursts
  (SUPERB), an ongoing pulsar and fast transient survey using the
  Parkes radio telescope. SUPERB involves real-time acceleration
  searches for pulsars and single-pulse searches for pulsars and fast
  radio bursts. We report on the observational setup, data analysis,
  multi-wavelength/messenger connections, survey sensitivities to
  pulsars and fast radio bursts and the impact of radio frequency
  interference. We further report on the first 10 pulsars discovered
  in the project. Among these is PSR~J1306$-$40, a millisecond pulsar
  in a binary system where it appears to be eclipsed for a large
  fraction of the orbit. PSR~J1421$-$4407 is another binary
  millisecond pulsar; its orbital period is $30.7$ days. This orbital
  period is in a range where only highly eccentric binaries are known,
  and expected by theory; despite this its orbit has an eccentricity
  of $10^{-5}$.

\end{abstract}


\section{Introduction}
In the past decade exploration of the high time resolution radio
Universe has begun to accelerate. This has resulted in numerous
discoveries with high scientific
impact~\citep{Hyman_2005,Kramer_2006,McLaughlin_2006,Hallinan_2007,Lorimer_2007,Osten_2008,Horesh_2015,Bannister_2016}. This
exploration is ever more tractable due to continuing technical
developments in telescope observing infrastructure and in computing
hardware and software. Some of the most exciting objects of study
necessitate real-time searches where the lag between the signal
being received by the telescope and being identified in a search algorithm
is reduced to the minimum possible; the reaction time typically needs
to be of the order of the event duration. Millisecond timescale
signals such as pulsars and fast radio bursts (FRBs) are thus quite technically challenging.

The High Time Resolution Universe South (HTRU-S, Keith et
al. 2010\nocite{Keith_2010}) survey has, between 2008 and 2014,
performed a Southern-sky search for pulsars and fast transients.
Amongst its pulsar discoveries HTRU-S identified the first ever
magnetar discovered in the radio \citep{Levin_2010}, the so-called
`diamond planet' \citep{Bailes_2011} and identified new high timing
precision pulsars \citep[see e.g.][]{Keith_2011}. Also due to its
frequency resolution, HTRU-S expanded the pulsar search parameter
space into regions of high dispersion measure (DM) and fast spin
periods \citep{Levin_2013}. The work of HTRU-S also confirmed the
existence of the cosmological population of FRBs
\citep{Thornton_2013}, initially signalled by the discovery of the
`Lorimer Burst' \citep{Lorimer_2007}.

Due primarily to limited computing resources in the past, the
discovery lag for pulsar and FRB signals has been anywhere from months
to several years. HTRU-S, like almost all previous pulsar and fast
transient surveys, was subject to this. However, with the advent of
fast networking capabilities between telescope hardware and
supercomputers, 
and the ubiquity of multi- and many-core
processors~\citep{Barsdell_2010}, it is now possible to process data
orders of magnitude faster. Applying these techniques provides an
improvement over the HTRU-S survey whereby new discoveries made with
the Parkes telescope can be acted upon in real-time. In the case of
FRBs, a real-time discovery enables the preservation of more
information about the burst and allows rapid action (or reaction) to
determine the source of the burst. This would help identify the many
basic properties of this population which remain unknown such as what
their all-sky/latitude dependent rate is, their spectra, their
brightness distribution and whether they are standard
candles. Real-time pulsar searches are equally essential but for a
very different reason. Due to the volume of data collected in pulsar
searches, long-term storage becomes a critical problem. With high data
rates and large surveys there is no time to search offline in order to
catch up. In the case of future telescopes such as the Square
Kilometre Array \citep{Braun_2015,Kramer_2015} offline searches will
not be possible as not all data will be recorded in long-term
storage. Real time pulsar searches can also open up new possibilities,
where one would want to take advantage of time-dependent
detectability. For example one could re-observe promptly if a pulsar
is boosted in flux density due to interstellar scintillation, if a
pulsar in a binary is found in a favourable part of the orbit, or if
an intermittent pulsar is emitting for only a small fraction of the
time.

In this paper we describe the SUrvey for Pulsars and Extragalactic
Radio Bursts (SUPERB) which aims to perform acceleration searches for
pulsars and single-pulse searches for FRBs (and pulsars) in real time,
bringing the discovery lag down to seconds. Furthermore, SUPERB
employs a network of multi-messenger telescopes working with the
Parkes Telescope, the primary telescope for the project. The search
for pulsars covers the widest acceleration range ever for a real-time
search and is thus in effect a demonstrator for what will be run on
next generation telescopes. In \S~\ref{sec:overview} we give an
overview of the project, the survey strategy and new software and
hardware innovations employed. \S~\ref{sec:infrastructure} describes
the data acquisition and pipeline processing performed, and
\S~\ref{sec:synergies} outlines the multi-messenger synergies with
other facilities across the electromagnetic spectrum and in other
windows. The first pulsar discoveries from the project are described
in \S~\ref{sec:psrs}, with particular focus on key interesting
individual objects. Finally we summarise in \S~\ref{sec:last}.

\section{Overview}\label{sec:overview}
In this section we describe the survey strategy for SUPERB.

With the suggested latitude dependence in the detectable FRB rate seen
in the HTRU Mid Latitude survey\citep{Petroff_2014}, albeit subject to
low-number statistics, it was decided to initially focus on a `high'
Galactic latitude region $15^{\circ} < |b| < 25^{\circ}$ (as indicated
by the previous analyses) to explore the cutoff of this effect. As the
survey progressed it was discovered that this latitude dependence
appears to become evident at even higher latitudes. The FRB rate seems
to increase above $|b| \sim 40-50^{\circ}$, apparently by as much as a
factor of $\sim3$, albeit with this also being subject to small number
statistics; this will be discussed further in the second paper in this
series by Bhandari et al., hereafter Paper 2. With this information it
was decided to increase the latitude range of the survey. A survey
extension over the initial sky region, dubbed SUPERBx\footnote{The
  SUPERB project is split across two project IDs in the Parkes data
  archive: P858 and P892. To obtain all the data, as described in the
  Data Access section at the end of this paper, one should query both
  of these project IDs.}, was undertaken to go all the way to the
Southern Galactic pole and, on the other side of the Galactic plane,
to $b=45^{\circ}$. In addition to these FRB-motivated selections in
Galactic latitude, sections through the Galactic plane previously not
covered to depths of 9-minute observations were included to
specifically search for pulsars which would have been missed by
previous studies (in particular HTRU-S). With the above selections in
Galactic latitude the longitude range was essentially set by the sky
visible at Parkes. Despite the limited time on sky available,
pointings in the Northern Celestial hemisphere were not excluded so as
to (a) maximise overlap with multi-wavelength facilities (see
\S~\ref{sec:synergies}), and (b) to enable future cross-calibration
with pulsar surveys running at other radio telescopes in the Northern
hemisphere.

Some of the region covered by SUPERB has been covered previously with
the same data-acquisition (but not data-processing, see
\S~\ref{sec:infrastructure}) setup, using 4.5-minute pointings in the
high-latitude component of the HTRU-S survey. The benefits of a
second-pass (or multiple passes) of the same area of sky are many when
it comes to pulsar searches: (i) intermittent pulsars which can be
`off' more often than not~\citep{Kramer_2006}, usually strongly
selected against, become detectable; (ii) when looking at high
Galactic latitudes in particular, there is more scintillation as
pulsar signals can be boosted in their apparent brightness due to
focusing in the turbulent interstellar medium~\citep{Rickett_1970};
and (iii) pulsars may be detected in parts of binary orbits where they
are not being eclipsed~\citep{Lyne_1990} or, for the most extremely
relativistic systems, in parts of the orbit where the acceleration is
within the search range. On top of these benefits one can leverage
real-time processing pipelines to realise that the optimal time to
re-observe a pulsar (or a pulsar candidate) is right now. Typical
processing lags in the past meant that attempted follow-up
observations days or weeks later were often unsuccessful requiring
repeated attempts to confirm pulsar candidates. Doing this correctly
and routinely also results in more efficient use of telescope
time. From the point of view of single-pulse searches, multiple passes
provide a longer time on sky increasing the likelihood that: (i) a
pulsar of any period exhibits a pulse at the bright end of its pulse
amplitude distribution; and (ii) a sufficient number of pulse periods
of a long-period pulsar occur during the observation. Long-period
pulsars are strongly selected against in both periodicity and
single-pulse searches, but the situation improves with observing
time. For FRBs that, in all but one case so far~\citep{Spitler_2016},
are not seen to repeat, these benefits do not apply. For these
sources, excepting the hinted-at latitude dependence, 
the current thinking is that it does not matter where we point so that
$N$ pointings of $M$-minute duration are just as good as performing a
single $N\times M$-minute pointing. Practically one loses time doing
the former as even the ideal case where there is no radio frequency
interference (RFI) and weather conditions are favourable one must
always slew between pointings, but it is only the former that allows a
sensible simultaneous pulsar search, except for a strategy where one
might stare at a globular cluster such as 47
Tucanae~\citep{Robinson_1995} or Terzan 5~\citep{Ransom_2005}.

\begin{figure*}
  \begin{center}
    \includegraphics[scale=0.32,trim=18mm 10mm 10mm 0mm, clip]{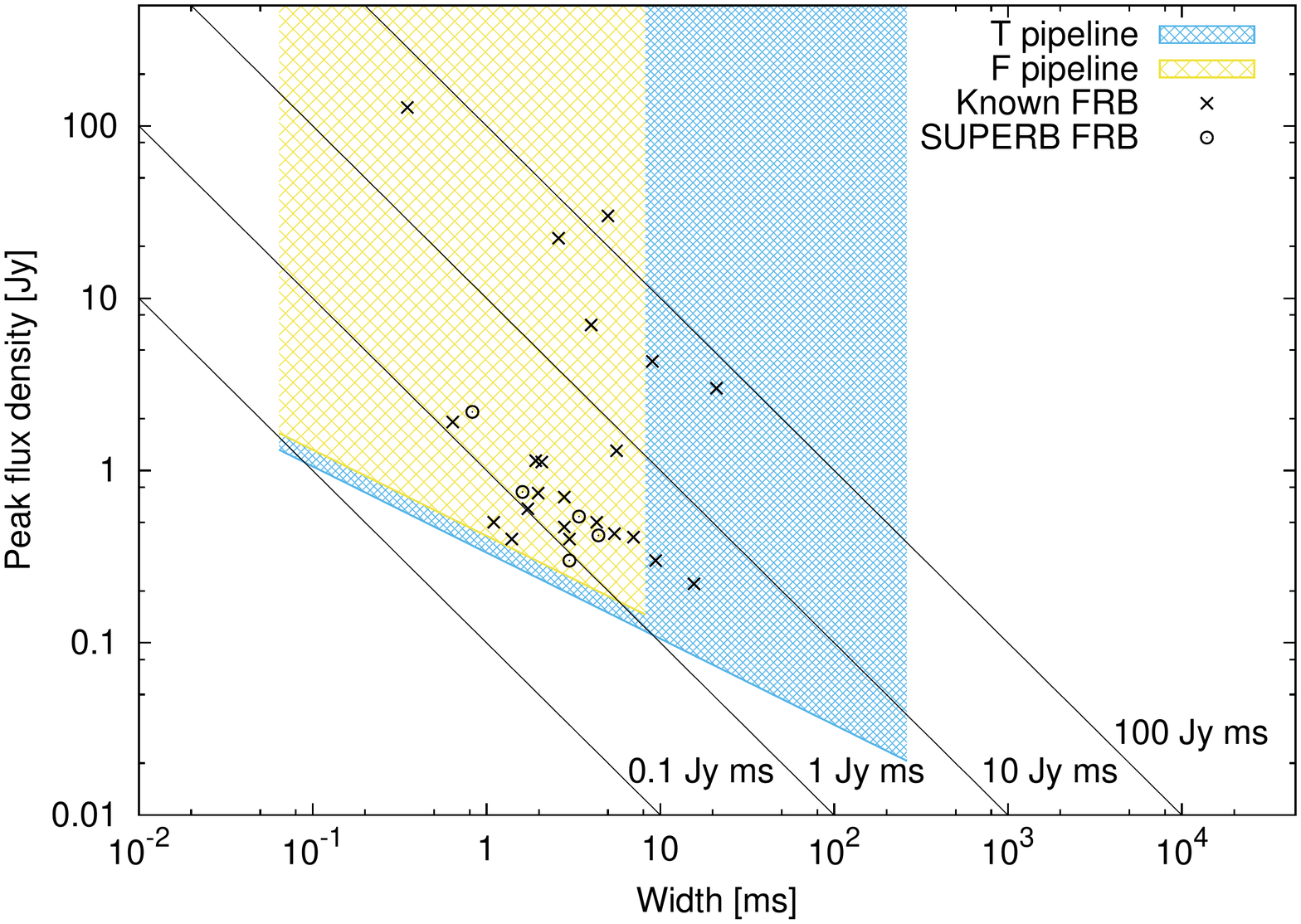}
    \includegraphics[scale=0.32,trim=18mm 10mm 10mm 0mm, clip]{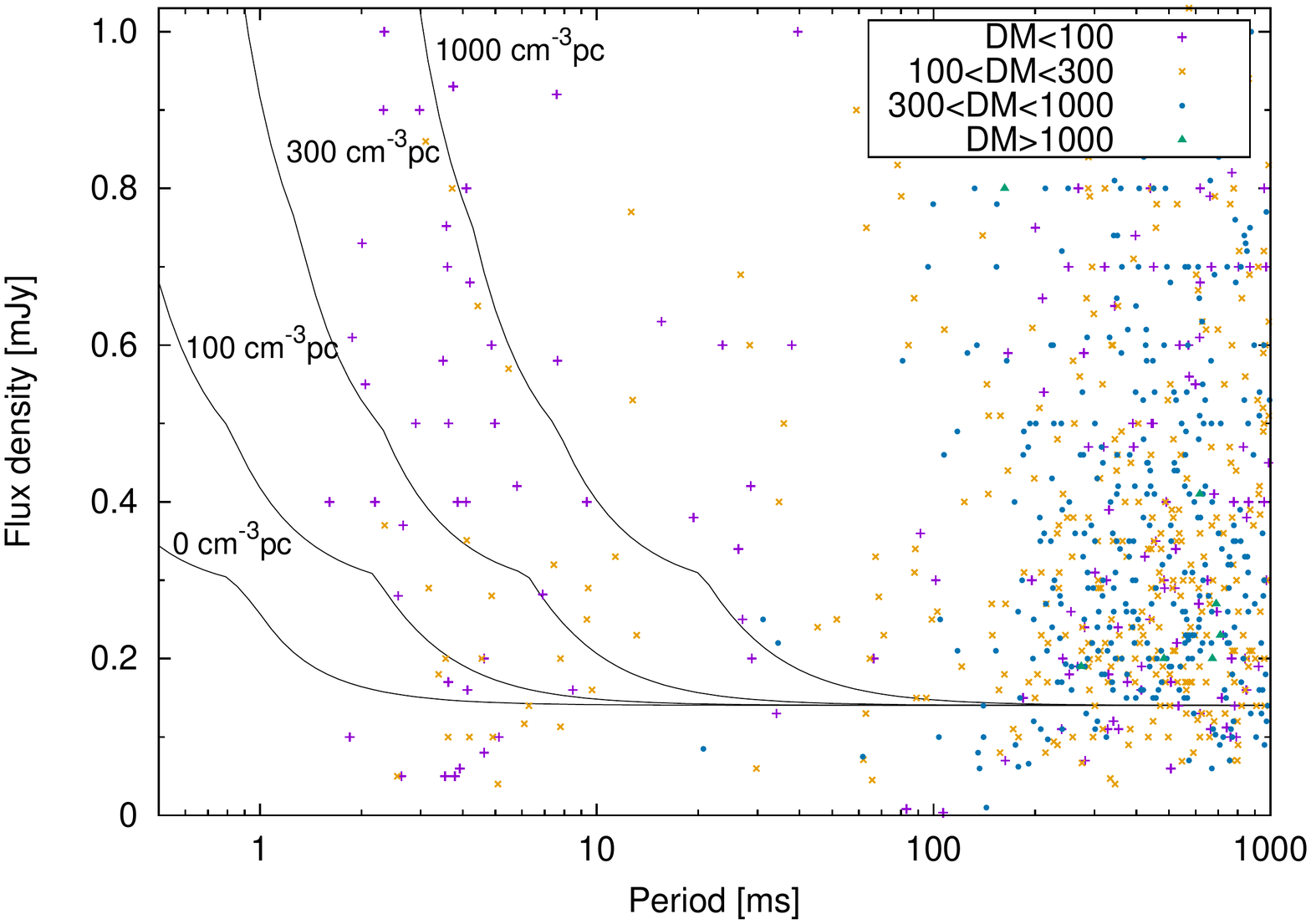}
    \caption{Left: Single-pulse search sensitivity plot for the SUPERB
      survey. Over-plotted are the 27 known FRBs with the relevant
      parameters published, including the five SUPERB discoveries (see
      the FRB Catalogue of Petroff et al. 2016, Keane et al. 2016 and
      Paper 2 for more details). Right: Periodicity search sensitivity
      plot for the SUPERB survey with the known pulsar population
      over-plotted in sub-groups (denoted by different symbols and
      colours) divided by their DM values.}\label{fig:sensitivity}
  \end{center}
\end{figure*}

With SUPERB we decided to perform 9-minute pointings, deeper than
previous HTRU-S observations which covered some of the SUPERB sky
region. Furthermore the SUPERB pointings are `in between' previous
pointings in two senses: (i) in tesselating the sky we first placed
the most sensitive central beam of the Parkes multi-beam receiver in
locations covered previously by the least sensitive outer-ring beams;
and (ii) we further offset the pointings by half of a half-power
beamwidth so that points previously at the half-power point were now
on-axis and vice-versa. These steps allow a more complete sampling of
the pulsar luminosity distribution. Repeating the same pointing
locations would repeat the incomplete coverage of the luminosity
function. In this way pulsars that fell into such `gaps' in previous
studies can now be detected. Figure~\ref{fig:sensitivity} shows the
sensitivity of SUPERB for both periodicity and single-pulse searches.

\begin{figure*}
  \begin{center}
    \includegraphics[scale=0.5, trim = 0mm 7mm 0mm 7mm, clip]{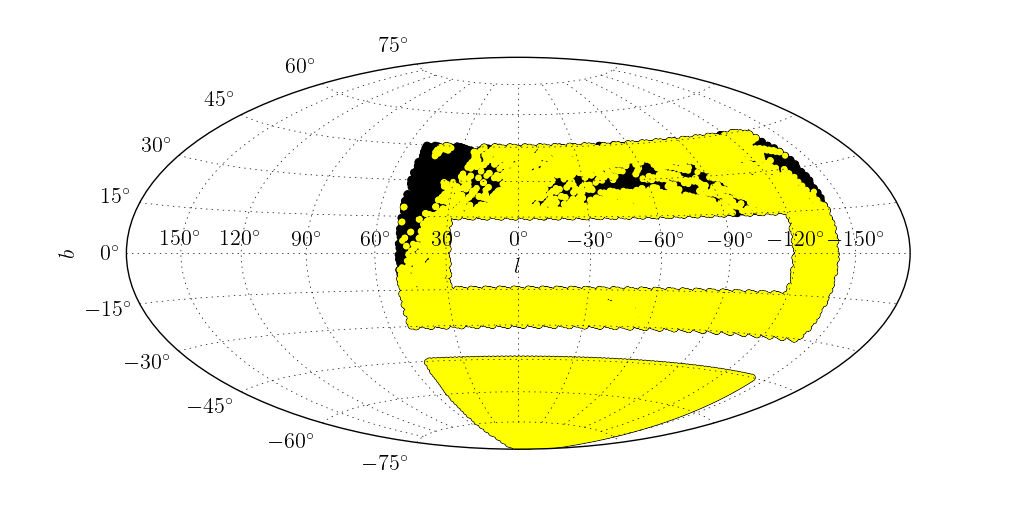}
    \caption{An Aitoff projection of the sky in Galactic
      coordinates. Those regions covered with 9-minute SUPERB
      pointings are marked in yellow. Pointings planned, but not yet
      observed before the end of January 2016 are marked in
      black.}\label{fig:pointings}
  \end{center}
\end{figure*}

The first SUPERB observing run was in April 2014 and lasted for 2
days. Subsequently it ran with some regularity from July 2014 to
January 2016. January 2016 to late 2016 saw a hiatus as Parkes was
used primarily to commission a phased-array-feed~\citep{Deng_2017},
but SUPERB resumed observations from December 2016. In this first
paper we consider the results up to the end of January 2016. The major
observing parameters are outlined in Table~\ref{tab:obs_params} and
the motivations for these choices are given below. Additionally, the
entire list of survey pointings performed to the end of January 2016,
illustrated in Figure~\ref{fig:pointings}, is included in the
additional online material associated with this article.

\begin{table}
  \centering
  \caption{Observational parameters of the survey. Where two values
    are given the first is for SUPERB (P858) and the second is for the
    SUPERBx (P892) components of the project.}
  \label{tab:obs_params}
  \begin{tabular}[h]{lcc}
    \hline  \hline
    Parameter & Value \\
    \hline
    Regions (P858) & $-120^{\circ} < l < 50^{\circ}$, $15^{\circ}<|b|<25^{\circ}$  \\
    & $29^{\circ} < l < 50^{\circ}$, $-25^{\circ}<b<25^{\circ}$ \\        
    Regions (P892) & $-140^{\circ} < l < 50^{\circ}$, $25^{\circ}<b<45^{\circ}$ \\
    & $-140^{\circ} < l < 50^{\circ}$, $-30^{\circ}<b<-25^{\circ}$ \\
    & $-140^{\circ} < l < 50^{\circ}$, $b<-45^{\circ}$ \\
    & $-140^{\circ} < l < -120^{\circ}$, $-25^{\circ}<b<25^{\circ}$\\
    $\tau_{\rm obs}$ (s) & $\sim560$ \\
    $N_{\rm beams}$ (planned) & 86424 and 180583\\
    $N_{\rm beams}$ (observed) & 71572 and 141512\\
    $T_{\rm samp}$ ($\upmu$s)& 64 \\
    $\Delta\nu$ (MHz) & 400 \\
    $\Delta\nu_{\rm chan}$ (kHz) & 390.625 \\
    $N_{\rm chans}$ & 1024 \\
    $N_{\rm samples}$ & $\sim$2$^{23}$ \\
    $N_{\rm bits}$ (online search) & 2 (periodicity), 8 (single pulse) \\
    $N_{\rm bits}$ (offline search) & 2 \\
    $N_{\rm bits}$ (archival) & 2 \\
    Archived data (TB) & 154 and 303 \\
    \hline
  \end{tabular}
\end{table}

\section{Infrastructure}\label{sec:infrastructure}
\begin{figure*}
  \begin{center}
    \includegraphics[scale=0.7,trim=0mm 0mm 0mm 0mm, clip]{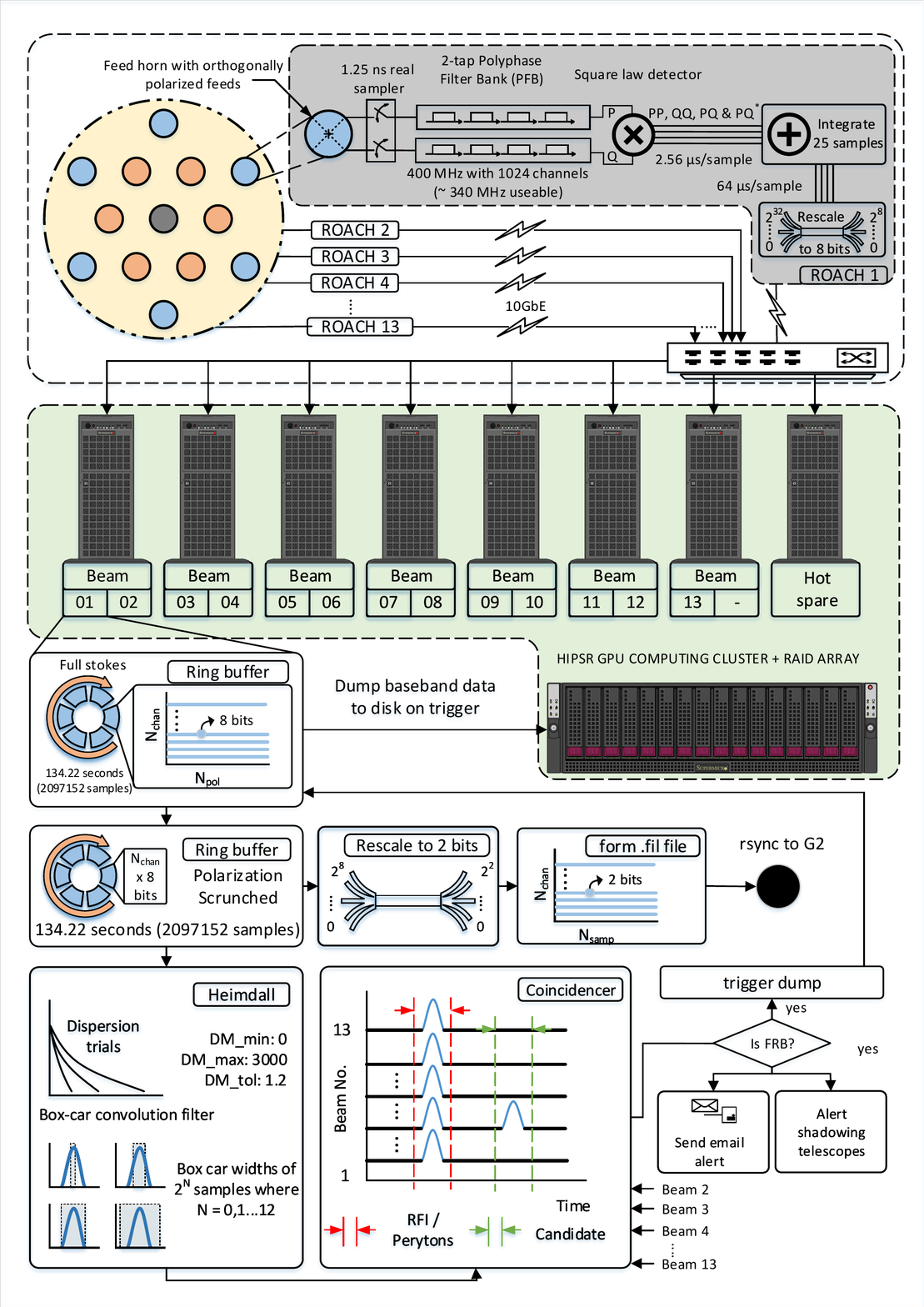}
    \caption{A schematic of the SUPERB acquisition and data analysis
      pipeline. The top most block shows the Parkes primary beam (yellow) with the 13-beam pattern of the multi-beam reciever over-plotted. Each beam is processed separately in the same way (grey block) on a Roach field programmable gate array --- Nyquist sampled in two orthogonal polarisations, formed into a filterbank, square-law detected to polarisation products and integrated down in time, rescaled from 32 bits to 8, then transferred to a GPU-enabled computing cluster for further searching (green block). The F pipeline single-pulse search happens on these machines --- a ring buffer in memory is searched for dispersed pulses satisfying several criteria (see main text) to identify these as FRB candidates. \textit{Bona fide} FRB candidate information is sent to other telescopes for data dumps (for shadowing telescopes) and/or follow-up.}\label{fig:vivek}
  \end{center}
\end{figure*}

One of the core objectives of SUPERB is to enable real-time pulsar and
FRB searches. Thus it requires comprehensive infrastructure with
minimal human input to allow for automated operation. 
Below we describe the data acquisition, processing pipelines and data
management scheme used in the project, which is shown schematically in
Figure~\ref{fig:vivek}.

\subsection{Data Acquisition}
SUPERB makes use of a modified version of the observing system
described by \citet{Keith_2010}. Here the Berkeley Parkes Swinburne
Recorder (BPSR) backend is used to digitise, filterbank, detect and
temporally average the signal from each beam of the Parkes 21-cm
multibeam receiver (Staveley-Smith et al. 1996). The output from BPSR
is an 8-bit, full-Stokes filterbank with 1024 frequency channels
spanning 400 MHz of bandwidth (1182 - 1582 MHz) and 64-$\upmu$s time
resolution. As noted by \citet{Keith_2010} the effective bandwidth is
$\sim$340 MHz due to the presence of strong RFI from communication
satellites emitting in the 1525-1559 MHz band
and roll-off at the band edges (see \S~\ref{sec:rfi}). The output from
BPSR is sent to the HI-Pulsar Signal Processor (HIPSR,
\citealt{Price_2016}) backend for further processing. It is here that
the observing systems of HTRU-S and SUPERB diverge. Where previously
data arriving in HIPSR would have been bit-compressed and the Stokes I
component written to disk and thence to magnetic tape, for the SUPERB
survey the data are instead pushed into a 120-s ring buffer. This ring
buffer serves two purposes; it provides input to a real-time transient
search and it enables full-Stokes data to be recorded upon receipt of
a trigger.
Following the transient search, the Stokes I component of the data is
bit-compressed to 2 bits per sample and is written to disk on
HIPSR. The removal of the magnetic tape storage step, employed by
HTRU-S and previous surveys, is significant as the writing of data to
tape had been the major contributing factor in discovery lag.

\subsection{Processing Hardware}
Upon completion of an observation at Parkes the data are streamed from
the HIPSR backend to the Green II (G2) supercomputing cluster located
at the Swinburne University of Technology. The G2 is composed of three
main components, the SwinSTAR and gSTAR compute clusters and a $\sim$5
PB lustre file system. For the purposes of SUPERB data processing we
are primarily interested in the number of GPUs available on each
cluster, to wit SwinSTAR provides 47 nodes each with two six-core
X5650 CPUs, 48 GB of RAM and two Nvidia Tesla C2070 GPUs and gSTAR
provides 61 nodes each with two eight-core E5-2660 CPUs, 64 GB RAM and
a Nvidia Tesla K10 GPU. SwinSTAR further provides 3 high-density
nodes, each with two six-core X5650 CPUs, 48~GB RAM and seven Tesla
M2090 GPUs. G2 uses a PBS-based queue system, employing Torque and
Moab for resource management and job scheduling, respectively.

\subsection{Processing Pipelines}
\begin{figure*}
  \begin{center}
    \includegraphics[scale=0.6,trim=0mm 0mm 0mm 0mm,clip]{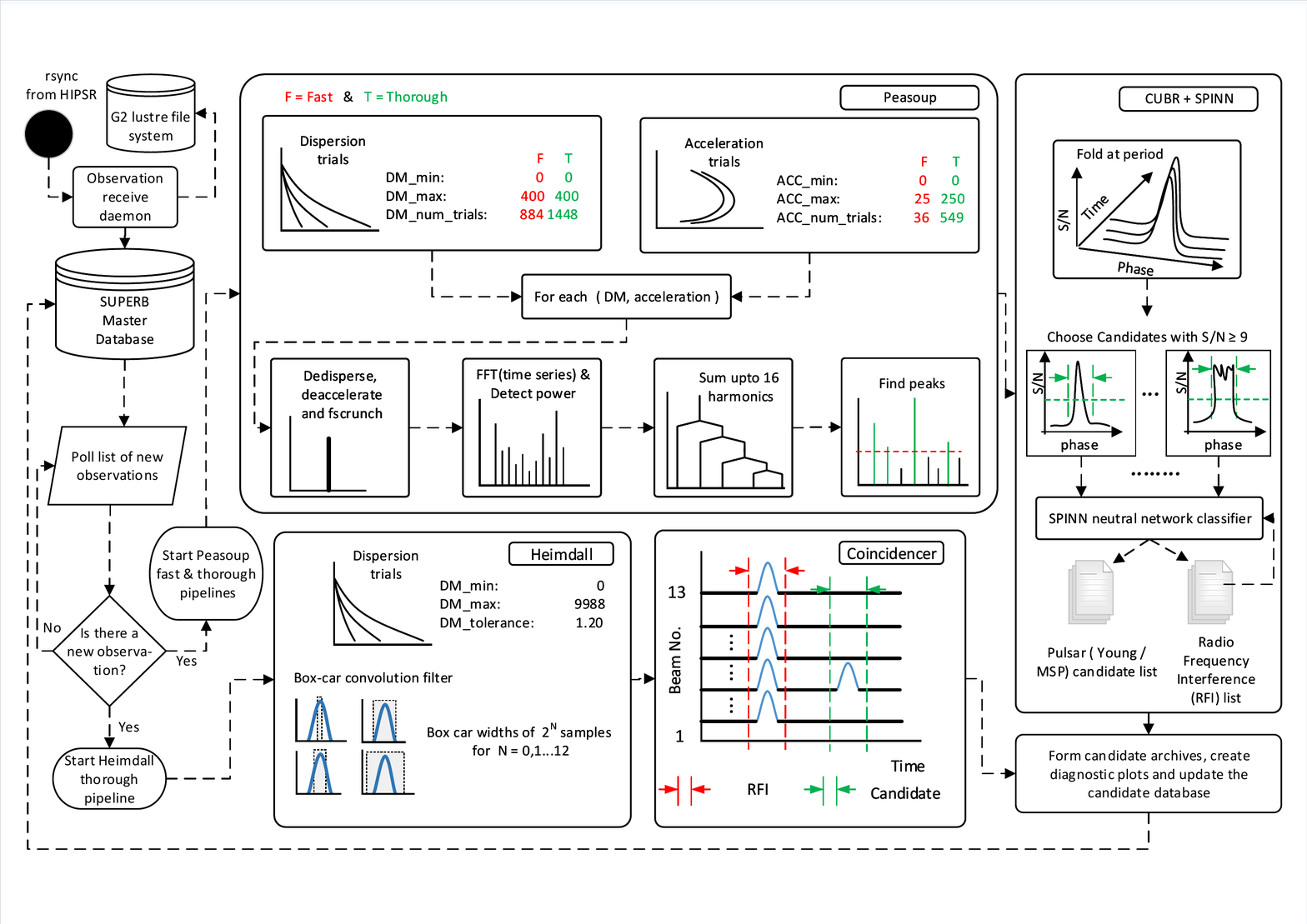}
    \vspace{-1cm}
    \caption{A schematic of the SUPERB processing pipelines on
      gSTAR.}\label{fig:vivek2}
  \end{center}
\end{figure*}

Data observed as part of SUPERB are searched for both periodic and
transient signals. In both cases the data go through both a fast (F)
and thorough (T) version of the respective processing pipeline. The
objective of the F pipeline processing is to enable real-time
processing, picking up the brighter signals with less extreme
properties, while the objective of the T pipeline processing is to
maximise our chances of discovery by searching a larger volume of
parameter space with higher resolution. Below we describe the F and T
versions of both our periodicity and transient searches, which is
shown schematically in Figure~\ref{fig:vivek2}.

\subsubsection{Transient Search Pipeline}
While the data are still stored in the 120-s ring buffer on the HIPSR
system they are searched using the F pipeline for single pulses while
still at 8-bit precision. This pipeline uses the
\textsc{heimdall}\footnote{http://sourceforge.net/projects/heimdall-astro/}
software package to search 16-s segments (or `gulps') of incoming
data\footnote{The size of these gulps is configurable; $16$~s is the
  default value used in SUPERB's pipelines.} over a range of pulse
widths, and dispersion measures for signals with properties resembling
those of real astrophysical pulses. The F pipeline search is
restricted to 1623 DM trials between 0 and 2000 pc cm$^{-3}$ and 13
width trials between $0.064$ and $262.144$~ms. The maximum DM was in
December 2015, increased to 3000 pc cm$^{-3}$. The search is run on
each beam individually to produce a list of candidates which are then
cross-correlated between beams to eliminate local RFI which occurs
uniformly, or in non-neighbouring beams. Frequency channels of known
sources of RFI are masked during the search to limit
contamination. Several cuts, described in the following section, are
applied to the resulting candidates for each gulp to search for FRBs;
the list of single-pulse candidates is not saved to disk.

After a pointing is completed the data are saved as filterbank files
and sent to the gSTAR supercomputer at Swinburne, they are searched
again for single pulses using the T pipeline. This pipeline is an
expanded version of the F pipeline on HIPSR which searches an entire
pointing over a larger parameter space to ensure that no viable
candidates are missed. For the T pipeline the pointing is searched up
to a maximum DM of 10,000 pc cm$^{-3}$ using 1986 DM trials, and
searched for pulses up to a maximum width of $262.144$~ms. The beams
are processed in a similar manner to the F pipeline, i.e. individually
and then compared to remove coincident RFI. Additional RFI excision
occurs in the form of frequency masking (as with the F pipeline) and
an eigenvector decomposition based algorithm~\citep{Kocz_2010}. The
list of single-pulse candidates are, in this case, saved to disk.

\begin{table*}
  \centering
  \caption{\small{The processing parameters for the F and T transient
      and periodicity search pipelines.}}
  \label{tab:transient_pipeline}
  \begin{tabular}[h]{lcc}
    Single-pulse pipeline parameter & F pipeline & T pipeline\\
    \hline
    DM range (pc cm$^{-3}$) & 0--2000 & 0--9988 \\
    DM trials, $N_{\rm DM}$ & 1623 & 1986 \\
    Width trials & $(1-2^{12})\times t_{\mathrm{samp}}$ & $(1-2^{12})\times t_{\mathrm{samp}}$ \\
    RFI excision methods & Bad channels & Bad channels \\
    & & eigenvector excision \\
    Periodicity search pipeline parameter & F pipeline & T pipeline\\
    \hline
    Maximum DM, $\rm{DM}_{\rm max}$ (pc cm$^{-3}$) & 400 & 400\\
    Trial DMs, $N_{\rm DM}$ & 884 & 1448 \\
    Maximum acceleration, $|a_{\rm max}|$ (m s$^{-2}$) & 25 & 250\\
    Acceleration trials, $N_{\rm acc}$ & 36 & 549 \\
    Number of harmonic folds performed, $N_{\mathrm{h}}$ & 4 & 4\\
    RFI excision methods & Bad channels & Bad channels \\
    & Birdie list & Birdie list\\
    & & Eigenvector excision\\
  \end{tabular}
\end{table*}

\subsubsection{Transient Candidate Selection}
After running the F and T pipelines the above parameters produce a
sizeable number of candidates. These are parsed according to the
following rules. In the F pipeline we apply:

\begin{gather}
  \mathrm{DM} \geq 1.5 \times \mathrm{DM}_\mathrm{Galaxy} \nonumber\\
  \mathrm{S/N} \geq 10 \nonumber\\
  N_\mathrm{beams,adj} \leq 4 \nonumber\\
  W \leq 8.192 \: \mathrm{ms} \nonumber\\
  N_\mathrm{events}(t_\mathrm{obs}-2\:\mathrm{s} \to t_\mathrm{obs}+2\:\mathrm{s}) \leq 5
  \label{eq:SP_Fpipeline}
\end{gather}

\noindent where DM and DM$_\mathrm{Galaxy}$ are the dispersion measure
of the candidate and the modeled DM contribution from the Milky Way
from the NE2001 model~\citep{cl02}, respectively, S/N is the
signal-to-noise ratio of the candidate, $N_\mathrm{beams,adj}$ is the
number of adjacent beams in which the candidate appears, $W$ is the
width of the candidate pulse, and the final cut describes the number
of candidates detected within a 4-second window centred on the time of
the candidate. If there are too many candidates in a time region
around the candidate of interest it is ignored, a precaution to reduce
the number of false positives due to RFI. The F pipeline only searches
for pulses from FRBs, hence the high-DM cutoff for viable
candidates. The excess DM factor of $1.5$ is arbitrary and might
conceivably result in missed FRBs in the F pipeline (or initial
classification of an FRB as a RRAT, see
\citealt{keane_2016}). 

\begin{figure*}
  \centering
  \includegraphics[width=2\columnwidth]{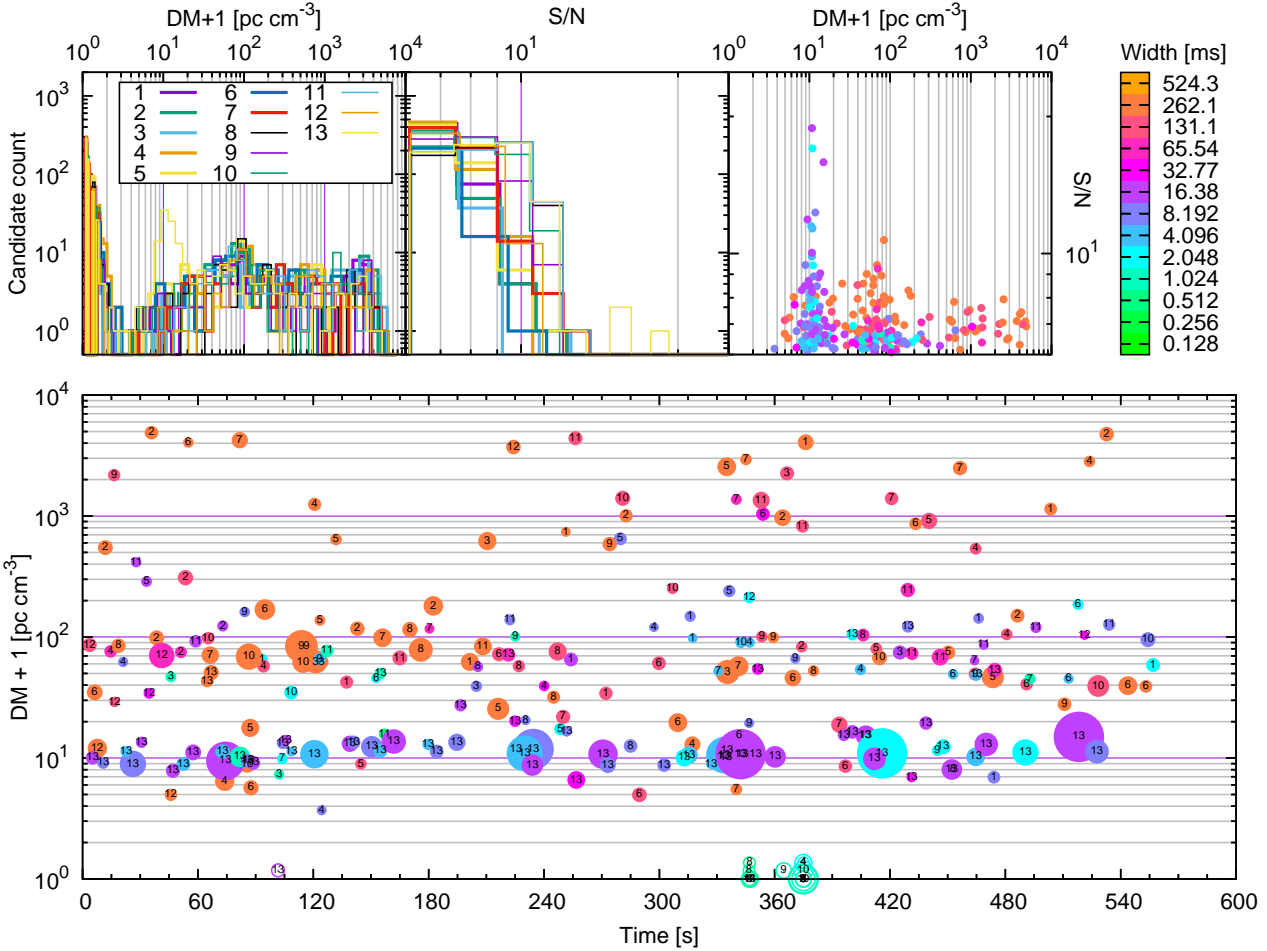}
  \caption{A single-pulse search candidate plot for a single SUPERB
    pointing. The top three panels show a histogram of events as a
    function of DM, then a histogram as a function of S/N, and a
    scatter plot of the DM and S/N of each candidate. The bottom plot
    shows the candidates in time and DM; here the colour corresponds
    to the pulse width and the number is the beam in which the
    candidate was detected. All candidates above a S/N of 8 are
    plotted here. In this example several pulses are evident in beam
    13 at a DM of approximately 10 pc cm$^{-3}$. We note that the
    bottom panel's axis label is DM$+1\;\mathrm{pc}\;\mathrm{cm}^{-3}$
    rather than simply DM. The reason for this is that a base-10
    logarithmic scaling is appropriate for our DM sampling, but there
    is still a need to display zero DM events to identify RFI. RFI
    signals peak at zero DM and depending on their time duration can
    be detected at higher DM values also. \label{fig:heimdall}}
\end{figure*}

In the T pipeline the data are searched for single pulses from pulsars
as well as pulses from FRBs to ensure that no candidates are missed in
the full processing of the data. For the T pipeline we apply:

\begin{gather}
  \mathrm{DM} \geq 2~\mathrm{pc}\;\mathrm{cm}^{-3} \nonumber\\
  \mathrm{S/N} \geq 8 \nonumber\\
  N_\mathrm{beams} \leq 4 \nonumber\\
  W \leq 262.144 \: \mathrm{ms} \nonumber\\
  \label{eq:SP_Tpipeline}
\end{gather}

\noindent where the parameters are as in
Equation~\ref{eq:SP_Fpipeline}. The T pipeline processing is intended
to detect lower S/N FRBs and single pulses from pulsars (see
Figure~\ref{fig:heimdall}).

\subsubsection{Periodicity Searching}
Periodicity searching of the SUPERB survey is performed using the
GPU-enabled pulsar searching code,
\textsc{peasoup}\footnote{\texttt{https://github.com/ewanbarr/peasoup}}. 
To implement a real-time pipeline, the three
high-density gSTAR nodes (21 Tesla M2070 GPUs) were reserved for all
SUPERB observing sessions. The search parameters are summarized in
Table \ref{tab:transient_pipeline}.

\begin{figure*}
  \centering
  \includegraphics[scale=0.6,angle=90,trim = 30mm 0mm 50mm 0mm, clip]{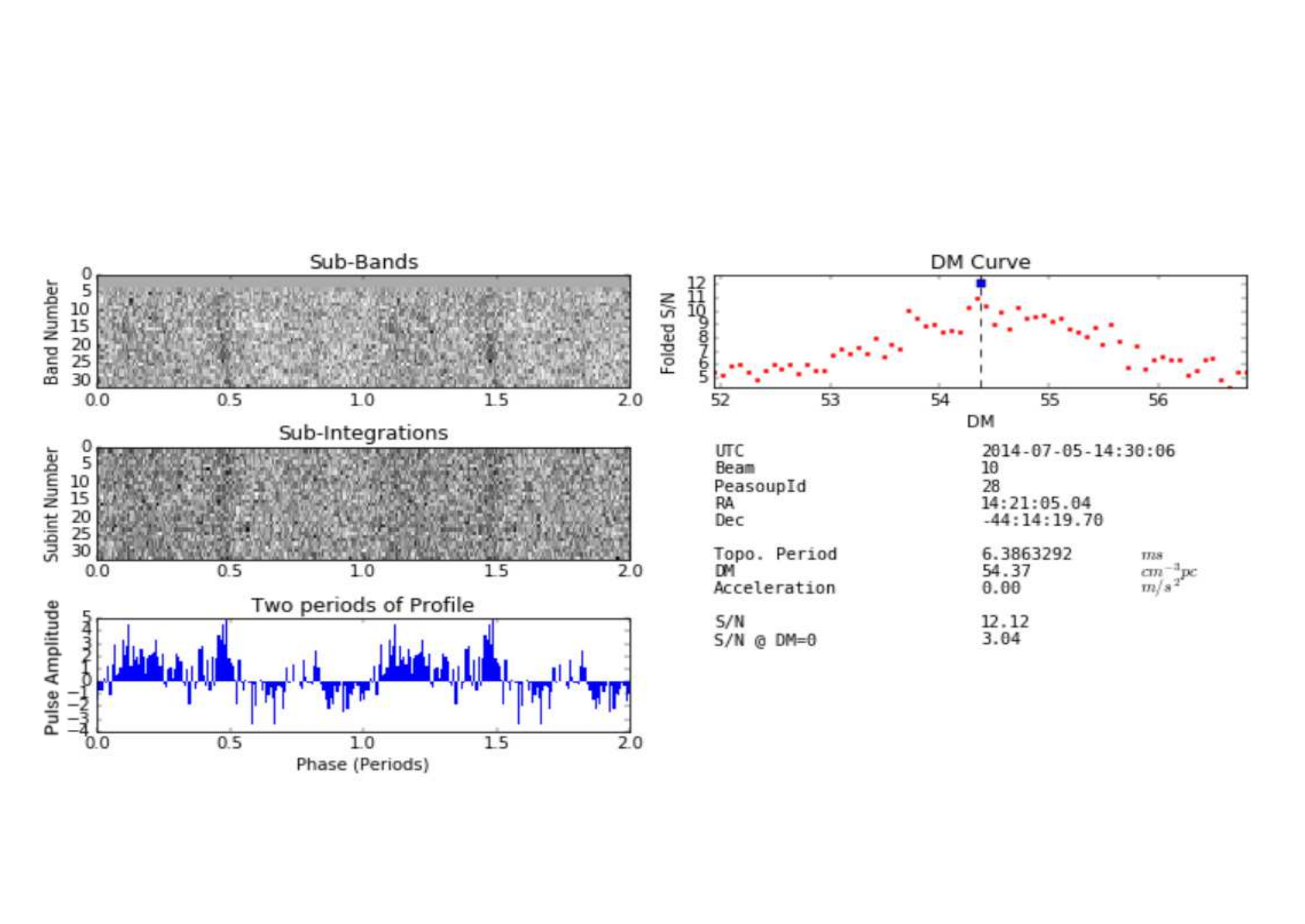}
  \caption{An example candidate plot from our pulsar searching
    pipeline: this is the first detection of PSR~J1421$-$44. The
    panels shown are, in grey scale, the period-folded signal as a
    function of frequency (top) and time (bottom), a folded pulse
    profile (blue profile) and a curve showing the S/N as a function
    of DM. Pertinent information about the pointing and candidate are
    also displayed. Diagnostic plots such as these are often used in
    examining candidates arising from our pipelines.}
  \label{fig:peasoup_cand}
\end{figure*}

\begin{figure*}
  \centering
  \includegraphics[scale=0.6,angle=90,trim = 30mm 0mm 50mm 0mm, clip]{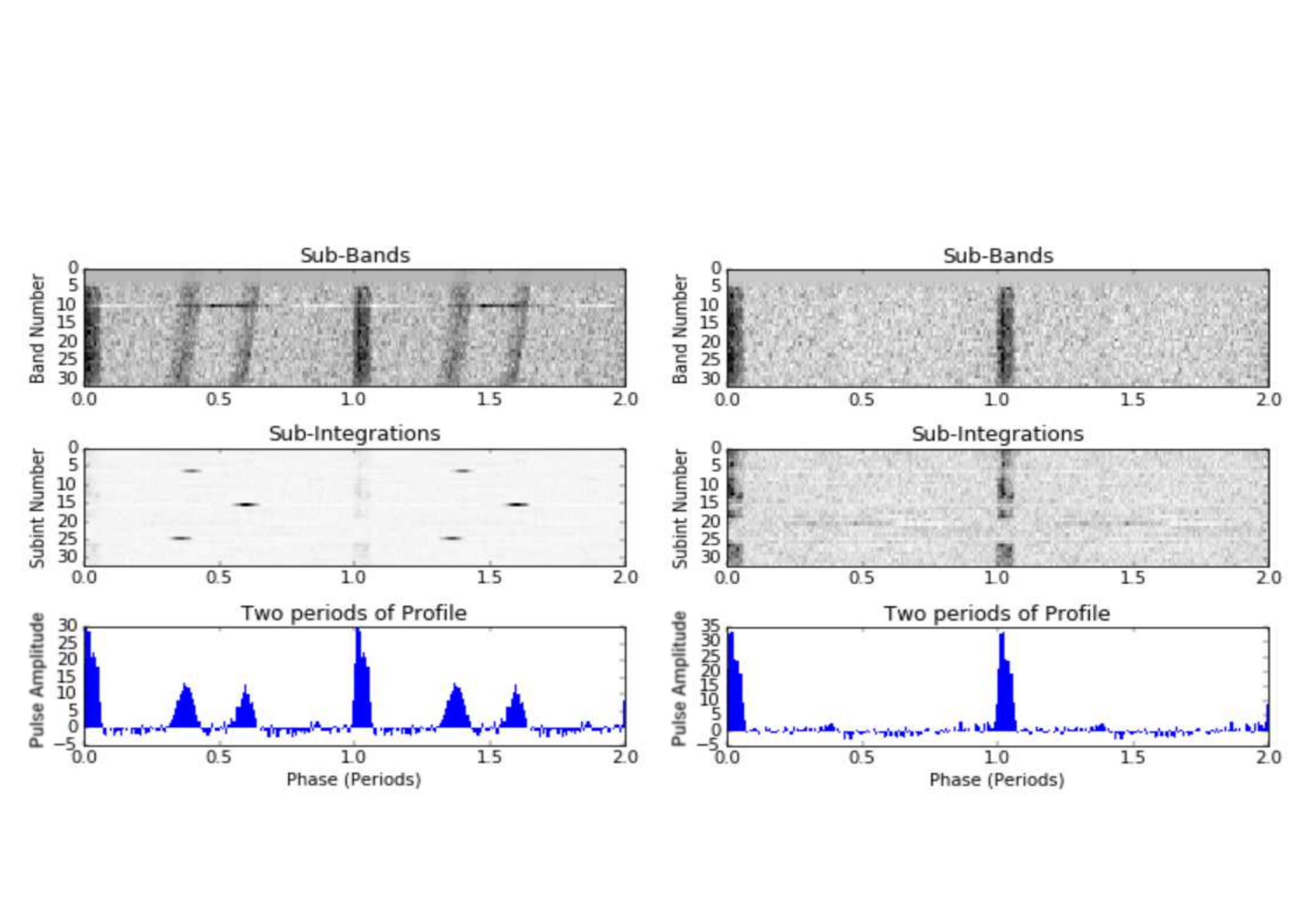}
  \caption{An observation of PSR J1759$-$1029 ($P=2.512$~s) before and
    after the application of the \textsc{cubr} RFI mitigation
    algorithms. The pulsar's signal is the vertical trail (two periods
    are shown for readability), and is demonstrably nulling in
    sub-integrations 20 to 25. Top left panel: a strong, periodic
    sine-wave like RFI is present in frequency band no. 10. Middle
    left panel: additionally, three bright dashes correspond to brief
    bursts of broadband interference; they are also visible in the
    sub-bands plot as curved trails, and generate secondary pulses in
    the overall folded profile (bottom panel). Right column: the same
    plots after the application of interference mitigation
    algorithms.}\label{fig:cubr_rfi}
\end{figure*}

In both F and T pipelines, we fold all candidates detected by
\textsc{peasoup} with a S/N higher than $9$, and always fold the $24$
brightest candidates of every beam. Figure~\ref{fig:peasoup_cand}
shows an example of candidate diagnostic plots from the periodicity
search. Motivated by the difficult RFI environment at Parkes and the
constraint of real-time processing for the F pipeline, a folding
software package named \textsc{cubr} has been written for the
survey. It can fold candidates in parallel, which reduces processing
time by a factor of $\sim5$ as compared to equivalent tools in the
\textsc{psrchive} package~\citep{hvm04}. It also has the ability to
delete interference signals directly in the folded data; abnormal
frequency channels or sub-integrations are identified using an outlier
detection method and the corresponding data are replaced by an
appropriately chosen constant value. This has the benefit of reducing
the difficulty of candidate evaluation, and in some cases identify
pulsars that would otherwise be entirely masked by interference (see
Figure~\ref{fig:cubr_rfi}).  \textsc{cubr}'s RFI mitigation algorithms
are described in length in \citet{vincent_thesis}.

\subsubsection{Periodicity Candidate Selection}
The combined output rate of both periodicity search pipelines is
approximately 4,000 candidates per observed hour, and they have
generated a total of 5.9 million folded candidates so far. Visual
inspection of all candidates is not a viable option considering the
thousands of hours of tedium it would involve and more importantly the
real-time discovery constraint we set for the F pipeline. We therefore
entirely transferred the task of candidate selection to a Machine
Learning algorithm. We use an improved version of the SPINN (the
Straightforward Pulsar Identification Neural Network) pulsar candidate
classifier~\citep{Morello_2014}. It is an artificial neural network
that evaluates candidates based on eight numerical features, and
outputs a score between 0 and 1 that can be interpreted as the
likelihood of being a pulsar.

\begin{figure}
  \centering
  \includegraphics[scale=0.35]{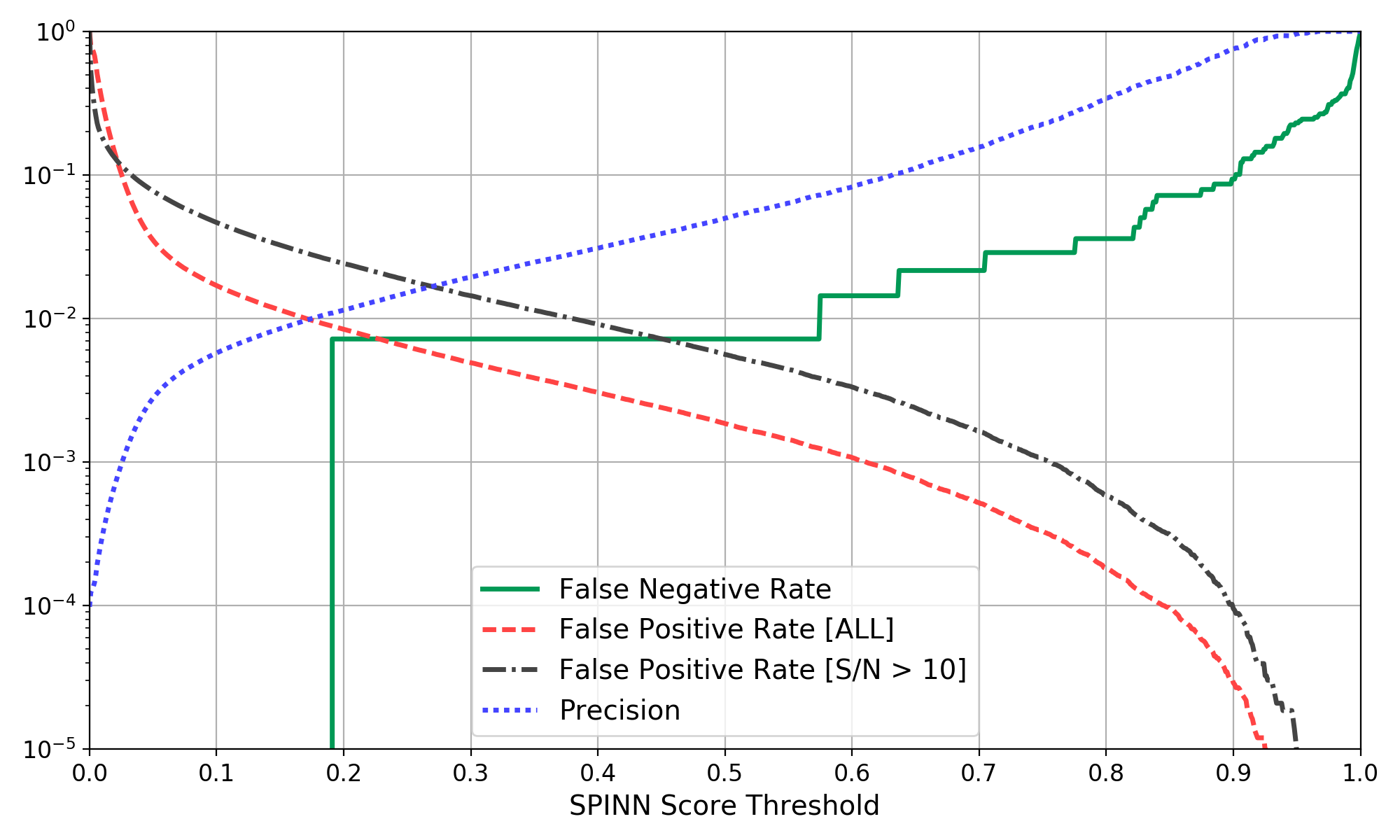}
  \caption{Classification error rates (log-scale) of SPINN as a function of output score selection threshold. The false negative rate is the fraction of pulsars missed (green line). The false positive rate is the fraction of spurious candidates incorrectly reported as pulsars (red line). We also evaluated it on the subset of candidates with $\mathrm{S/N} > 10$ (black line), since those cannot be trivially rejected by the classifier based on their lack of statistical significance. Precision (blue line), is the fraction of genuine pulsars contained in the sample of candidates selected by the classifier, a relevant metric for the real-time detection pipeline.}
  \label{fig:metrics_vs_score}
\end{figure}

SPINN was trained on a large sample of candidates obtained by running
our search and folding pipeline on the HTRU-S Intermediate Latitude
survey~\citep{Keith_2010}, which was observed between 2008 and 2010
with the same telescope, instrument, and sampling and integration
times, but covered an area of the sky that does not overlap with that
of SUPERB. We could therefore rigorously test the classification
accuracy of SPINN on a sample of SUPERB candidates, knowing that none
of those could have been `seen' by the algorithm during
training. Using the ephemerides published in the ATNF pulsar
catalogue~\citep{mhth05}, we manually identified every known pulsar
detection found by \textsc{peasoup} during the first part of the
survey (up to April 2015). This gave us a test sample of 139 pulsar
detections along with 1,418,598 non-pulsar candidates which were
scored by our classifier. From there we computed classification error
rates as a function of score selection threshold; the results are
summarized in Figure~\ref{fig:metrics_vs_score}.

We use different neural network score thresholds for the F and T
pipeline. For real time discovery, we select candidates scoring higher
than 0.85 to ensure a small false positive rate ($\approx$ 1 in
10,000) so that discovery alerts are reliable and can be acted upon
quickly. In the T pipeline, we value search completeness most highly
and therefore tolerate a higher false positive rate at the cost of a
longer round of visual candidate inspection; we typically inspected
candidates down to a score of 0.5, and 0.4 for millisecond pulsar
candidates. This involved looking at on order of 10,000 candidate
plots over the course of the entire survey.




\subsubsection{Peryton Pipeline}
At the inception of the project we planned to search for
`perytons'~\citep{Burke_Spolaor_2011a}. These transient signals had
been detected in archival data from Parkes with more than a decade of
discovery lag and, although clearly terrestrial in nature, their
source had yet to be pinpointed as of 2014 when SUPERB was
beginning. We searched for perytons as for transient events above but
with two slight modifications. Firstly, to take advantange of the fact
that perytons are local and therefore detectable in most or all 13
beams of the receiver, the 13 beams are added to produce a new
filterbank. Astrophysical events, which are typically in a single
beam, are thus suppressed in S/N by a factor of $\sqrt{13} \sim 3.6$
whereas peryton signals are boosted by up to this factor in the
resultant data set. This data is then searched as per the single-pulse
search of a single beam. In identifying peryton candidates the
number-of-beams sifting rule does not apply, and no DM cut is
applied. SUPERB thus has the ability to discover peryton events in
real time and did this when they first occurred during SUPERB
observations in January 2015. The combination of the ability to
identify these signals without any discovery lag, and the data from
the RFI monitor installed at Parkes in December 2014 allowed the
source of the perytons to be identified. The simultaneous coverage up
to 3~GHz allowed us to identify the carrier frequency of $\sim
2.4-2.5$~GHz and through a process of deduction the culprit:
unshielded microwave ovens on site. This work is discussed in more
detail in \citet{Petroff_2015b}. Since this time a live peryton search
is not routinely performed but the publicly available survey data
still contains these signals should others wish to pursue this study.

\subsubsection{Future Pipelines}
\textit{(i) Fast Folding Search}: Due to short observing lengths the
survey suffers a sensitivity loss to long-period pulsars that cannot
be detected through their single pulses. It has long been known that
the Fast Folding Algorithm (FFA, \citealt{Staelin_1969}) offers the
possibility to recover sensitivity to these pulsars, even in short and
noisy observations. Due to its computationally intensive nature, the
FFA has seen only sparse use in blind pulsar surveys to date
\cite[although it has seen use in targeted searches, see
  e.g.][]{Kondratiev_2009}. However, renewed interest in
FFAs~\citep{Cameron_2017} and their implementation on many-core
compute architectures has led to the development of new codes that can
be applied to the SUPERB data set. The application of an FFA to SUPERB
data will address known biases in our processing and greatly improve
our capability to discover pulsars at the long-period extreme of the
population. This pipeline commenced in April 2017 and the results of
this will be reported at a later date.

\textit{(ii) Low-Level RFI Mitigation Algorithms}: In both the
transient and periodicity search pipelines, most of the burden of RFI
rejection is currently placed on the final candidate selection
stage. This approach is not optimal since interference occurring
during a pulsar observation can be strong enough to make the pulsar's
signal completely unrecognizable at the candidate inspection stage,
even by a highly trained expert. In particular, a very common and
unwelcome occurrence in SUPERB data is that of bright, broadband
non-dispersed pulses lasting several milliseconds that are
simultaneously visible in all beams of the receiver. These negatively
impact the detectability of slow pulsars in the Fourier domain and are
a prolific source of false positives to the single-pulse
search. Existing tools such as the \textit{rfifind} routine of the
\textsc{presto} package~\citep{Ransom_2002} do not effectively
mitigate their effects. An attractive approach here is the application
of spatial filtering~\citep{Leshem_2000,Raza_2002,Kocz_2010}, which is
particularly efficient at identifying and canceling any interfering
signal present in a large number of beams; while normally applied to
baseband data, we are currently investigating the use of spatial
filtering on incoherent filterbanks, allowing its deployment on
archival data.


\section{Multi-wavelength Synergies}\label{sec:synergies}
The Parkes telescope is the primary instrument of the SUPERB
project. However it is joined in its search efforts for varying
amounts of time by a number of additional facilities, some of which
work simultaneously and some of which react to triggers from
Parkes. In this section we describe the shadowing (simultaneous
observations) and triggering performed in the electromagnetic spectrum
as well as multi-messenger searches for counterparts to Parkes
discoveries.

\subsection{Shadowing}
\begin{figure*}
  \begin{center}
    \includegraphics[scale=0.5]{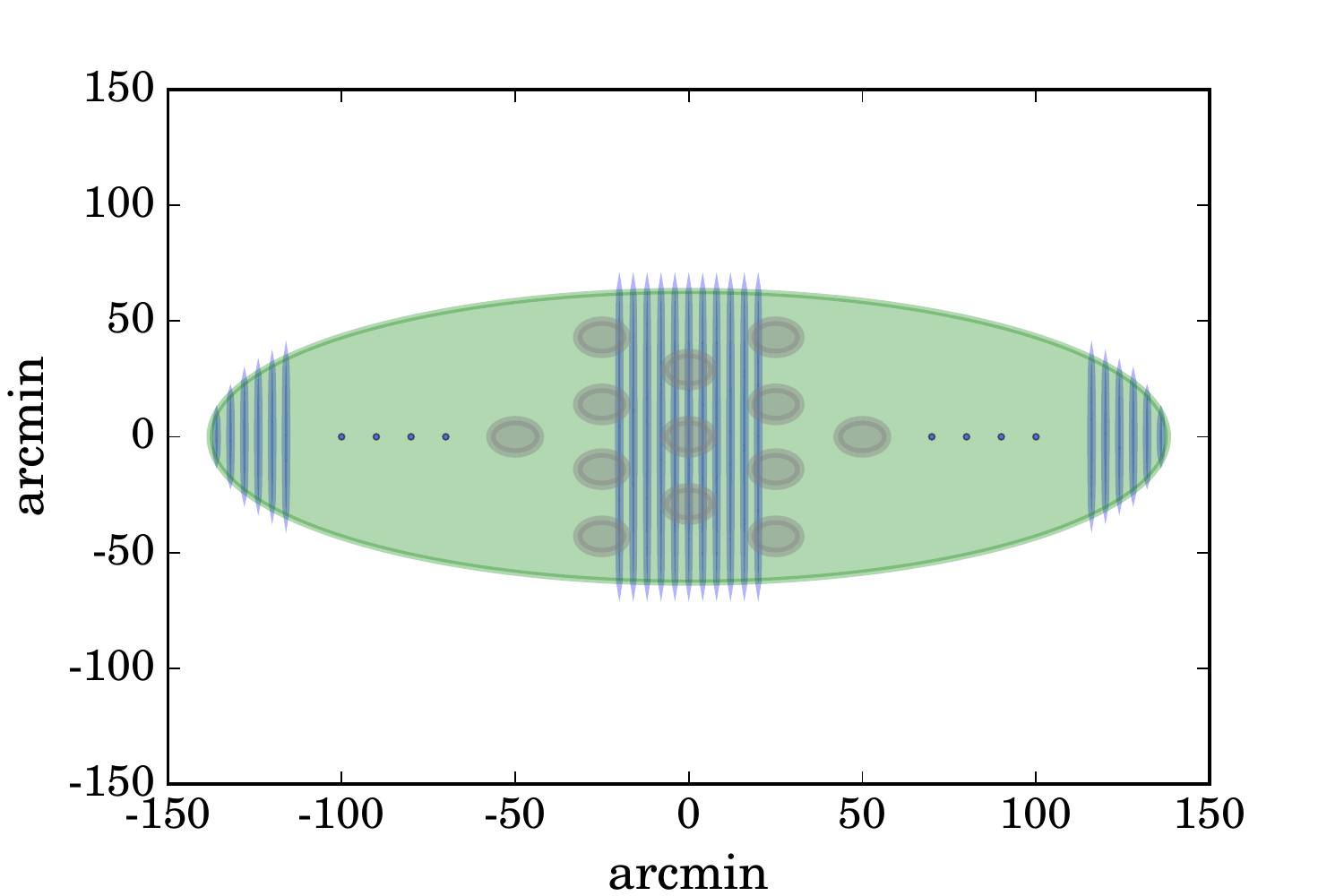}
    \includegraphics[scale=0.5]{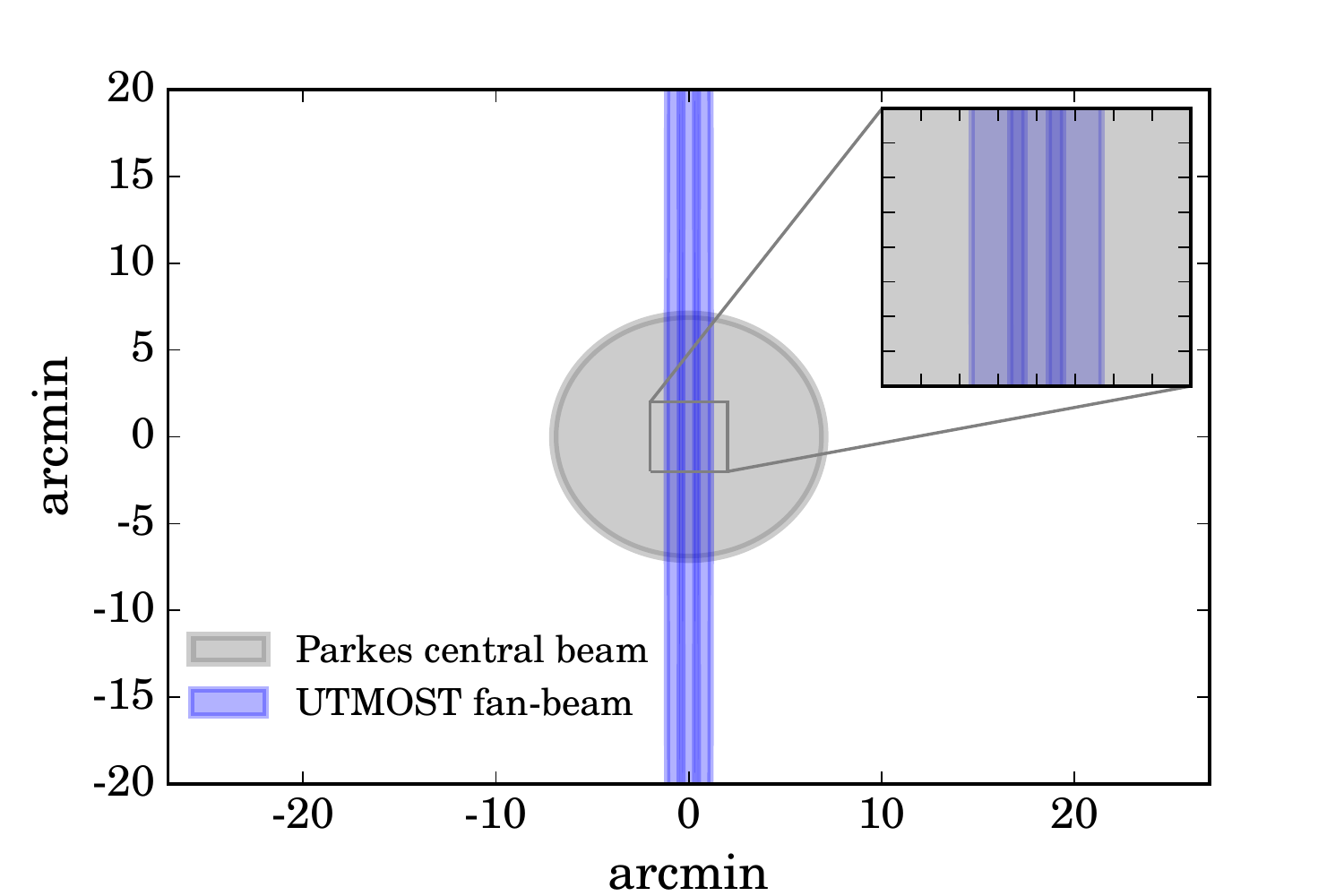}
    \caption{The left-hand panel shows a Molonglo primary beam (green)
      at 843 MHz over-laid on the 13-beam sky footprint of Parkes
      (grey). Some Molonglo fan-beams are also shown. 
      A simultaneous detection at both telescopes can thus be used for
      more precise localisation than can be achieved at either
      telescope operating by itself. The right-hand panel shows a
      zoom-in on a single Parkes beam and shows three overlapping
      Molonglo fan-beams. Parkes FRBs are typically (although not
      always) in a single beam; Molonglo FRB detections are typically
      in 1 to 3 fan-beams. Combining this information an FRB can be
      localised an order of magnitude more precisely than with Parkes
      alone.}\label{fig:localise}
  \end{center}
\end{figure*}

The observations at Parkes are shadowed to various degrees by other
telescopes. At the time of writing, this has been done with the
upgraded Molonglo Synthesis Telescope (UTMOST), the Giant Metrewave
Radio Telescope (GMRT) and the Murchison Widefield Array (MWA).
At the conception of the SUPERB project in late 2013 the intention was
to shadow Parkes with Molonglo at all times. 
By the start of the first SUPERB observing session in April 2014 the
shadowing infrastructure was in place, but the Molonglo upgrade,
was still in progress. 
Upon detection of an FRB a ring buffer of the polarisation power data
at Parkes is dumped and a signal is sent to Molonglo to dump the
single polarisation complex voltage data from every module in the
array. The idea is to detect the signal at both telescopes and, in the
case of an FRB localise the signal an order of magnitude more
precisely than either telescope can do by itself and to obtain vital
information on spectra. This concept is illustrated in
Figure~\ref{fig:localise}.

\begin{figure}
  \begin{center}
    \includegraphics[scale=0.4, trim=0mm 20mm 0mm 20mm, clip ]{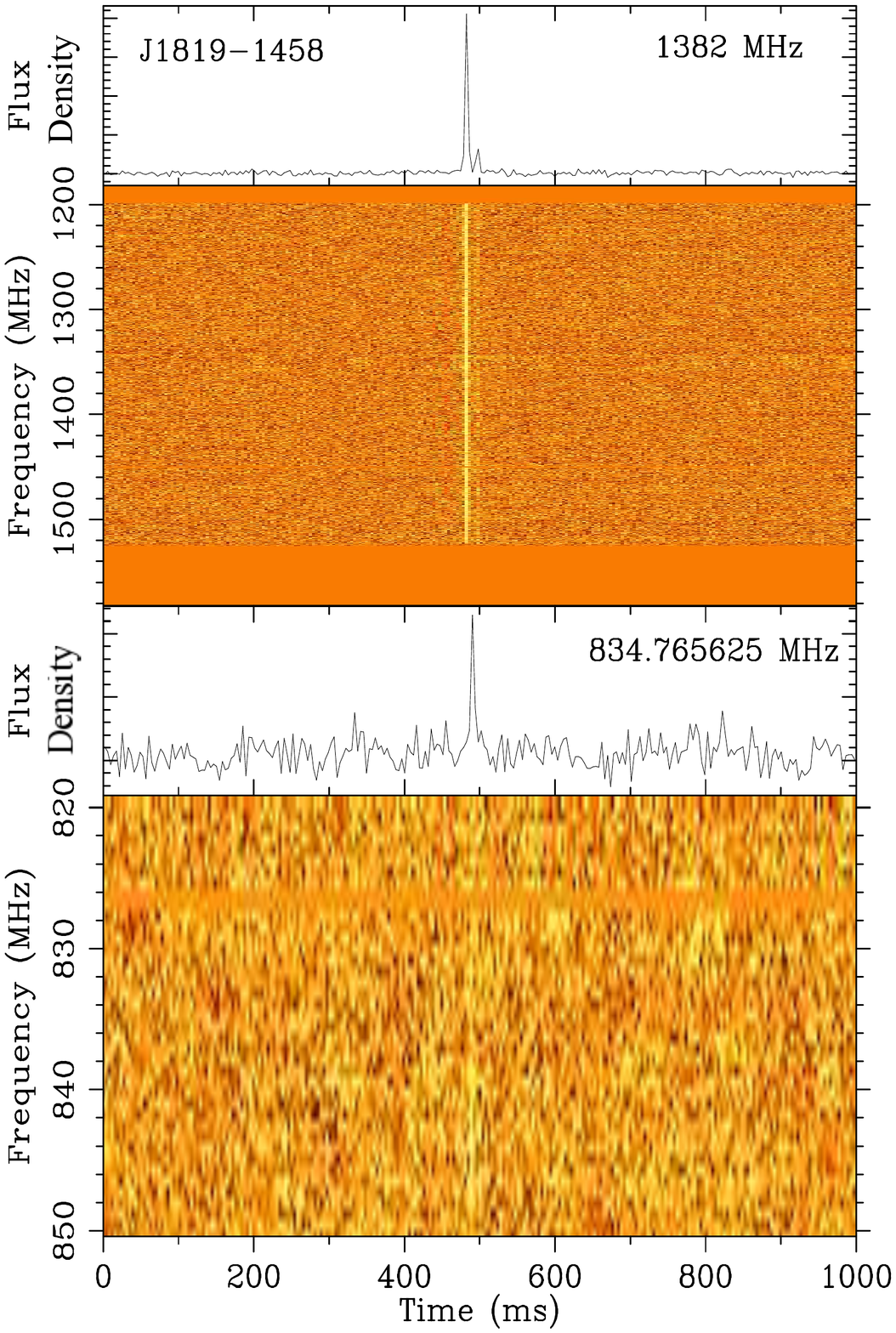}
    \caption{First simultaneous detection of a pulse from J1819$-$1458 at Parkes and Molonglo. The top panel shows the detection at Parkes, and the bottom the simultaneous detection at Molonglo. Note the different vertical scales for frequency.}\label{fig:1819}
  \end{center}
\end{figure}

Tests of this joint observing mode were performed using the erratic
pulsar J1819$-$1458 \citep{McLaughlin_2006} which has a known spectrum
and exhibits very bright pulses every $\sim$minute at 1.4 GHz at
Parkes. 
The initial tests revealed that the sensitivity of Molonglo was less
than $1\%$ of its final target sensitivity and as such 
Molonglo shadowing, though available, was not utilised for most 
of the observations reported here. However in the interim the Molonglo
upgrade has proceeded at pace, further tests with J1819$-$1458 were a
resounding success (see Figure~\ref{fig:1819}), and the sensitivity is
now at a level above $10\%$ of the theoretical optimum. This progress
is well illustrated by the recent independent discoveries (i.e. not in
tandem with Parkes) of 3 FRBs by UTMOST~\citep{Caleb_2017}. 

The MWA ($\sim 600$ ${\rm deg}^2$ field of view at $200$~MHz) is also
used, on occasion, to shadow SUPERB.
Occasional shadowing has been performed since mid-2015 with more
routine shadowing since January 2016.
Data are recorded using the standard mode of the MWA's hybrid
correlator, recording visibilities at a cadence of 500 ms and 40 kHz.
Furthermore, since the MWA's sensitivity is a strong function of
zenith angle, in order to keep the loss in sensitivity minimal and
ensure quality calibration, Parkes pointings that are west-most are
preferentially selected while the MWA is available for shadowing (but
see \S~\ref{sec:rfi} for discussion of the RFI implications of
this). The large field of view
means there are no addition requirements in terms of additional
calibration observations as multiple suitable sources are typically
present for any given MWA pointing.

The 30-antenna GMRT is also used, on occasion, to shadow SUPERB,
in particular in the 325-MHz band, where the 84$^{\prime}$ (FWHM) beam
is well-matched to Parkes as it fully encompasses all 13 beams of the
multi-beam receiver, albeit with non-uniform sensitivity.
Simultaneously detecting a potential low-frequency counterpart allows
one to constrain both the spectral nature and scattering properties of
any FRBs in addition to the ability to precisely localise it in the
sky (at the level of a few arcseconds). Shadowing with GMRT is complex
as one must consider the down time due to the difference in slew
rates, the need for approximately hourly phase calibration
observations.
Data are recorded using the high-time resolution mode of the GMRT
software correlator~\citep{Roy_2010} whereby the visibilities are
recorded once every 125 ms, at a spectral resolution of 65 kHz, over a
bandwidth of 16.66 MHz, centred at a frequency of 325.83 MHz. Despite
the inevitable temporal and dispersive smearing expected for any
potential counterparts to the FRB signals, this still ensures good
detection prospects; e.g. a putative low frequency counterpart of FRB
110220 would be detectable as a 7$\sigma$ event. The time allocation
and coordination considerations typically allow shadowing about 10\%
of the SUPERB survey. The common visibility is ensured by
preferentially going for Northerly pointings ($\delta > -40^{\circ}$)
that are past transit for Parkes during the times the GMRT is used
(again see \S~\ref{sec:rfi}). 

\subsection{Triggering}
When an FRB is found in the F Pipeline burst search an alert is issued
to the observers via email which can be visually inspected and
assessed. If the signal is judged to be a FRB detection, a trigger is
issued to collaborators for multi-wavelength follow-up. SUPERB
maintains agreements with a large number of telescopes and
collaborations to search for the signatures of FRBs across the
electromagnetic spectrum. At the highest energies, SUPERB triggers the
High Energy Spectroscopic System (HESS, \citealt{hess}) operating in
the range 10 GeV -- 10 TeV. At X-ray wavelengths the \textit{Swift}
satellite~\citep{swift} is triggered, which then observes X-ray
photons from $0.2-10$~keV.

Additionally, triggers are sent to the 2-m Liverpool telescope in La
Palma, the 1.35-m Skymapper telescope in New South Wales, Australia,
the 1-m Zadko telescope in Western Australia, the 2.4-m Thai
telescope, the 8.2-m Subaru telescope in Hawaii and the 10-m Keck
telescopes in Hawaii, the 6.5-m Magellan telescopes in Chile, and the
Blanco 4-m telescope in Chile using the Dark Energy Camera (DECam). At
radio wavelengths, triggers are sent to the 64-m Sardinia radio
telescope capable of observing at 1.4~GHz, the GMRT, and the
MWA~\citep{mwa}. Internally, the SUPERB collaboration also operates
the Australia Telescope Compact Array (ATCA) to image the field of the
FRB at 5.5 and 7.5~GHz. Follow-up timing of pulsar discoveries is also
performed by some of the radio telescopes in this network.

\begin{table}
  \centering
  \caption{The network of instruments alerted to SUPERB FRB triggers.}\label{tab:multiscope}
  \begin{tabular}{lc}
    \hline \hline
    Telescope name & Band/filters \\
    \hline
    H.E.S.S. & 10 GeV -- 10 TeV \\
    \textit{Swift} & 0.2 -- 10 keV \\
    Liverpool Telescope & $R$\\
    Skymapper Telescope & H$_\alpha$, $ugvriz$\\
    Zadko Telescope & $R$ \\
    Thai National Observatory & $R$ \\
    Blanco Telescope & $ir$VR \\
    Subaru Telescope & $r'i'$ \\
    Keck Telescope &  $400-1100$~nm \\
    Magellan Telescope & J \\
    MWA & 185~MHz \\
    GMRT & 1.4~GHz, 610~MHz \\
    Sardinia Radio Telescope & 1.4~GHz \\
    Effelsberg & 1.4 GHz \\
    ATCA & $4-8$~GHz \\
  \end{tabular}
\end{table}

\subsection{Multi-messenger Searches}
The SUPERB project has agreements in place with multi-messenger
facilities to search for counterparts to FRB events. A subset of FRB
progenitor models involve merger events of compact objects and as such
may have an associated gravitational wave counterpart. Furthermore the
redshift ranges for many of the FRBs are plausibly within the relevant
redshift horizon for ground-based gravitational wave detectors. As
such we have an agreement in place with the LIGO consortium to
identify counterparts in the Advanced LIGO dataset.
Some FRB progenitors may also exhibit a neutrino signal. Earth-based
neutrino detectors, which are sensitive to muon interactions with
neutrinos, use the Earth itself as a filter against background muon
signals, essentially look through the planet. 
We have an agreement with the ANTARES collaboration~\citep{antares} to
search for neutrino signals associated with FRBs. 

\subsection{Public Alerts}
From 2018-04-01 SUPERB will issue public alerts of FRB discoveries,
with an associated Astronomer's Telegram. The alert will be in the
VOEvent Standard in the FRB format currently being finalised.
While at present Parkes is the dominant FRB search machine, having
discovered $22$ of the $31$ FRBs known, it is envisioned that other
instruments, in particular CHIME, ASKAP and MeerKAT, will soon
contribute significantly to the known population. As such it makes
sense to create and adopt a world standard for FRB followups, and it
is widely considered that public alerts are the best way to do
this. The lead time to change to public alerts, as opposed to
immediate adoption, is (a) to allow us to satisfy our commitments
under agreements with partner instruments; and (b) allow finalisation
of the format for FRB VOEvents and development of associated tools,
which is currently underway.
%
%
%
%
%
%
%

\section{Results}\label{sec:psrs}
In this section we describe the first results from the survey,
including verification of the expected sensitivity, the impact of
radio frequency interference on the survey, and discoveries of FRBs
and pulsars.

\subsection{Survey Sensitivity Verification}
To verify that the expected sensitivity of the survey is being
achieved we keep track of all of the detections of previously known
pulsars, detected in the pipelines described above.

\textit{Single-pulse search}: The sensitivity of the SUPERB survey to
single pulses from pulsars was compared to the sensitivity from the
HTRU survey~\citep{BurkeSpolaor_2011} for a subset of the detected
pulsars. The theoretical flux denity of a single pulse detected by the
multi-beam receiver is given by the modified radiometer equation:

\begin{equation}\label{eq:SPradiometer}
  S_\mathrm{peak} = \frac{(\mathrm{S/N}) T_\mathrm{sys} \beta}{G \sqrt{n_\mathrm{p} \, \Delta \nu \, W}} 
\end{equation}

\noindent where $S_\mathrm{peak}$ is the peak flux density of the
pulse, S/N is the signal-to-noise ratio as before, $\beta$ is a
correction factor to account for small loses due to the digitisation
($\beta \approx 1.066$ for 2-bit digitisation in our case), $G$ is the
gain of the telescope beam, $n_\mathrm{p}$ is the number of
polarisations summed to create the signal, $\Delta \nu$ is the
bandwidth, and $W$ is the pulse width. For a single pulse detected in
the primary beam of the receiver with a S/N of 10 and a width of 1 ms
this corresponds to a peak flux density of 0.5 Jy.
The sensitivity to single pulses was found to be unchanged between the
SUPERB and HTRU surveys and consistent with flux densities expected
from Equation~\ref{eq:SPradiometer}.

\begin{figure}
  \centering
  \includegraphics[scale=0.45]{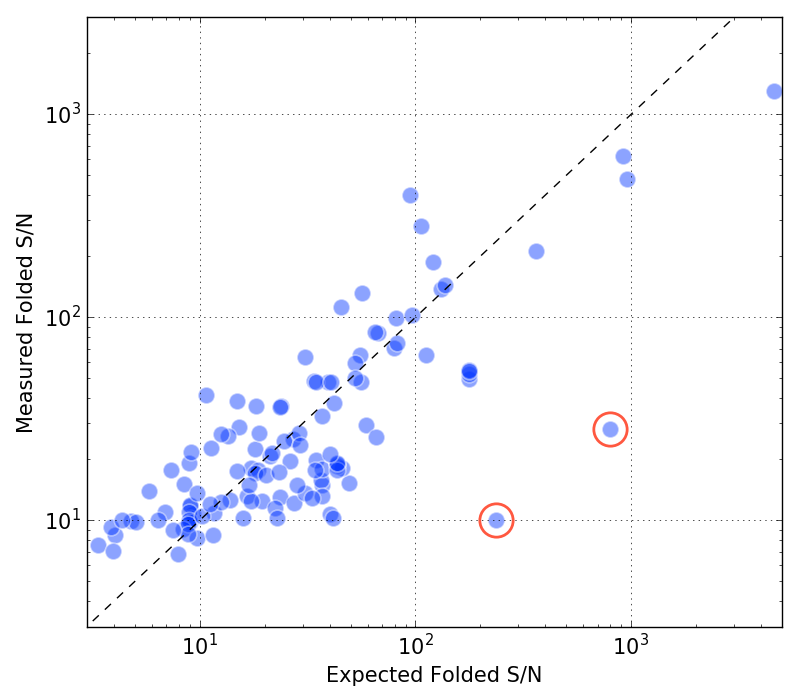}
  \caption{Measured versus expected folded S/N for known pulsars whose
    position was observed during the first part of the survey
    (excludes SUPERBx). The dashed line materializes the expected 1:1
    correlation. Expected S/N has been computed under the assumption
    of a 5\% pulsar duty cycle. The outliers circled in orange
    correspond to two detections of J0904$-$7459, whose previously
    catalogued flux density at 1400 MHz appears to have been vastly
    larger than what our data indicates (see text).}
  \label{fig:expected_observed_snr_folded}
\end{figure}

\textit{Periodicity search}: We directly folded the survey data (up to
and including February 2015) using ephemerides of all known pulsars
that had a published mean flux density, $S_{\mathrm{mean}}$, at 1400
MHz and whose position was observed at least once. After visually
inspecting the output we identified 124 detections of such pulsars and
recorded their S/N. Using the modified radiometer equation appropriate
for folded observations one can compute the expected folded S/N of a
pulsar with duty cycle $\delta$:
\begin{equation}\label{eq:expected_folded_snr}
  \mathrm{S/N} = \frac{g S_{\mathrm{mean}} G \sqrt{n_\mathrm{p} \,
      \Delta \nu \, T_{\mathrm{obs}}} }{ \beta T_{\mathrm{sys}}
  }\sqrt{\frac{1-\delta}{\delta}}
\end{equation}
where $ T_{\mathrm{obs}}$ is the integration time, and
$g=\exp(-\alpha^2 / 2\alpha_0^2)$ is an adjustment to the boresight
gain due to positional offset of the pulsar --- the beam response is
Gaussian; $\alpha_0 = 6.0$ arcmin for the Parkes multi-beam
receiver. Figure \ref{fig:expected_observed_snr_folded} displays the
measured vs. expected folded S/N for our sample of pulsars. The vast
majority of sources follow the identity line as it should be. The only
notable exception is PSR J0904$-$7459: it was seen twice with a S/N
approximately 30 times lower than expected, and failed to be detected
in two other observations where it should have been seen with S/N
$\approx$ 150. These repeated and consistent discrepancies suggest
that the catalogued flux density of B0904$-$74 (J0904$-$7459) is
erroneously high, an inference which is supported by a recent large
scale study of pulsar spectral properties~\citep{Jankowski_2017}. As a
result of this realisation the pulsar catalogue (version 1.56) has now
been updated to reflect our observations (R.N. Manchester,
priv. comm.).

We also checked which known pulsars of our sample were not properly
detected by our search pipeline. The notable non-detections of
\textsc{peasoup} are listed in Table
\ref{tab:notable_non_detections}. None of those are surprising as they
can be explained by either the presence of RFI, or the fact that the
FFT tends to lose sensitivity to signals with long periods. SPINN gave
a score lower than 0.5 only to a single pulsar observation, whose
candidate plot was heavily affected by impulsive RFI. This can also be
seen in Figure \ref{fig:metrics_vs_score}, where the false negative
rate at a score of 0.5 is not quite zero.

\begin{table}
  \centering
  \caption{List of notable non-detections of \textsc{peasoup}. We also
    ran the \textsc{seek} routine of the well-established
    \textsc{sigproc} pulsar searching package, and obtained three
    faint detections. No major discrepancy can be seen though. One can
    also note the potential benefits of an FFA-based pulsar
    search. Note that J1910$+$0714 was affected by impulsive RFI, and
    J1105$-$4357 was impacted by the presence of periodic RFI at a
    period of 1000.0~ms.}\label{tab:notable_non_detections}
  \begin{tabular}{lcccc}
    \hline \hline
    Name  &  P (ms)    & DM     & Folded S/N & \textsc{seek} S/N\\
    \hline
    J1105$-$4357  &  351.1     & 38.3   & 12.2       & 6.9 \\
    J1842$+$0257  &  3088.3    & 148.1  & 13.9       & 6.7 \\
    J0633$-$2015  &  3253.2    & 90.7   & 14.0       & $-$ \\
    J0636$-$4549  &  1984.6    & 26.3   & 14.0       & $-$ \\
    J1846$-$7403  &  4878.8    & 97.0   & 14.1       & 6.4 \\
    J1910$+$0714  &  2712.4    & 124.1  & 17.6       & $-$ \\
    J1945$-$0040  &  1045.6    & 59.7   & 24.6       & $-$ \\
  \end{tabular}
\end{table}

\subsection{FRBs}
The first FRB discovered by SUPERB is FRB~150418. This source has been
reported in \citet{Keane_etal} and further discussed in many susequent
publications as we now recap. FRB~150418 is at low Galactic latitude
($b=-3.3\degree$) but despite this is clearly extragalactic with
$\mathrm{DM}/\mathrm{DM}_{\mathrm{Milky\;Way}}$ of $4.2$ ($2.4$) according to the NE2001
(YMW16) model of the electron density in the
Galaxy~\citep{cl02,ymw16}. In brief the main point of subsequent
discussion has been the statistical association (and its implications)
discussed in \citet{Keane_etal}. This association, with a source in
the elliptical galaxy WISE J071634.59$-$190039.2, was based on its
contemporaneous brightening and an estimate of the likelihood of its
light curve ($\sim 99\%$ probability of association, based on 5
observation epochs). Further observations at the highest angular
resolutions show that the source (which must be more compact than
$\sim 10$~pc) is most likely an
AGN~\citep{Bassa_2016,Giroletti_2016}. As more data became available
the light curve of the variable radio source became ever better
characterised and the statistical significance of the association
reduced (to $\sim 92\%$, based on 24 epochs, see e.g.
\citealt{WilliamsBerger,Johnston_2017}). Unless a repeat FRB is seen
from the source this association is thus likely to remain somewhat
controversial, at least until such time as the statistics of longer
($\sim$day) timescale variability at below $100$~$\mathrm{\upmu Jy}$
becomes clearer.

Four further FRB discoveries from SUPERB will be reported in detail,
along with their multi-wavelength and multi-messenger followup, in
Paper 2 in this series.

\subsection{New Pulsars}
\begin{figure*}
  \begin{center}
    \vspace{-6cm}
    \includegraphics[scale=0.8]{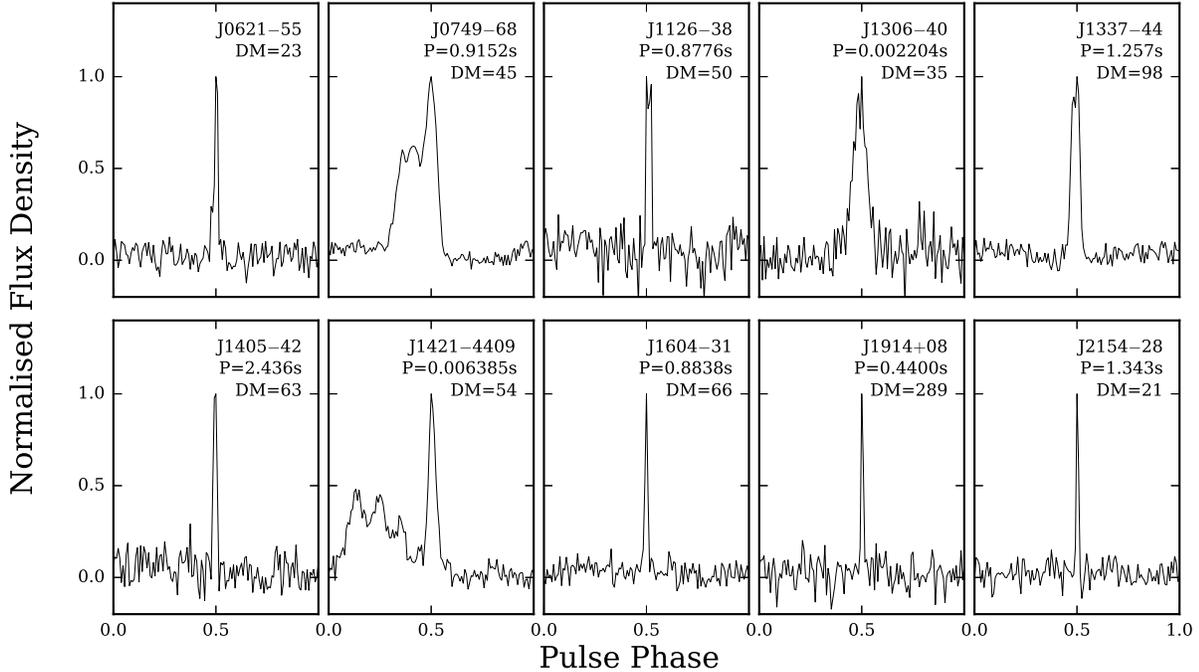}
    \vspace{-6cm}
    \caption{{Pulse profiles of the first 10 pulsar discoveries from
        SUPERB. Each panel shows one pulse period and the pulsar name
        is given in the top right corner, along with its rotation
        period in seconds and dispersion measure in units of
        $\mathrm{cm}^{-3}\,\mathrm{pc}$. J0621$-$55 
        does not yet have a determined
        periodicity.}}\label{fig:pulsars}
  \end{center}
\end{figure*}

The first 10 pulsars discovered in this survey are listed in
Table~\ref{tab:pulsars}. For seven of these the positions listed
denote the phase centre of the beam in which they were discovered and
should only be taken as indicative prior to a full timing solution
being obtained; the exceptions are identified
below. Figure~\ref{fig:pulsars} shows the pulse profiles.

\begin{table*}
  \setlength{\tabcolsep}{0.13cm}
  \caption{The parameters of 10 newly-discovered pulsars from the SUPERB Survey. The timing solution for PSR~J1421$-$4409 is detailed in Table~\ref{tab:1421_par}. We list the positions in both Equatorial coordinates with uncertainties, and the equivalent Galactic coordinates omitting those same uncertainties, the spin period ($P$), the DM and the NE2001-derived distances of these pulsars. Values in parentheses are the nominal 1-$\sigma$ uncertainties in the last digits. We understand that two of these pulsars have been independently identified, but not yet published, in two other ongoing surveys denoted here as: * = GBNCC, ** = PALFA.}  
  \label{tab:pulsars}
  \begin{tabular}{lcccccccc}
    \hline \hline
    PSR~name & R.A. (J2000)  & Dec. (J2000)   & $l$ & $b$ & $P$ & DM & Dist & Comment \\
    & (h:m:s) & $(^\circ:\,':\,'')$ & $(^{\circ})$ & $(^{\circ})$ & (ms) & (cm$^{-3}$pc) & (kpc) \\
    \hline
    J0621$-$55 & 06:20.7(5) & $-$56:05(7) & 264.822 & $-$26.416 & - & 22 & 1.1 & RRAT \\
    J0749$-$68 & 07:50:50(1) & $-$68:44:27(4) & 281.013 & $-$20.110 & 915.171299(2) & 26 & 1.1 & scintillates \\
    J1126$-$38* & 11:26.3(5) & $-$38:38(7) & 285.230 & 21.286 & 887.55(1) & 46 & 1.7 & \\
    J1306$-$40 & 13:06:56.0(5) & $-$40:35:23(7) & 306.108 & 22.186 & 2.20453(2) & 35 & 1.2 & MSP, intermittent \\
    J1337$-$44 & 13:37.1(5) & $-$44:43(7) & 311.412 & 17.386 & 1257.52(9) & 96 & 3.5 & nuller \\
    J1405$-$42 & 14:05.8(5) & $-$42:33(7) & 317.249 & 18.233 & 2346.80(4) & 64 & 2.0 & - \\
    J1421$-$4409 & 14:21:20.9646(3) & $-$44:09:04.541(4) & 319.497 & 15.809 & 6.38572883816(3) & 54.6 & 1.6 & MSP, binary \\ 
    J1604$-$31 & 16:04.4(5) & $-$31:39(7) & 344.118 & 15.380 & 883.883(5) & 63 & 1.9 & - \\
    J1914$+$08** & 19:14.3(5) & $-$08:45(7) & 43.327 &  $-$1.042 & 440.048(2) & 285 & 7.0 & - \\
    J2154$-$28 & 21:54.8(5) & $-$28:08(7) & 20.854 & $-$51.089 & 1343.35(2) & 28 & 1.2 & - \\
    \hline
  \end{tabular}
  \end{table*}

\begin{figure}
  \begin{center}
    \includegraphics[scale=0.4, trim=0mm 0mm 0mm 18mm, clip]{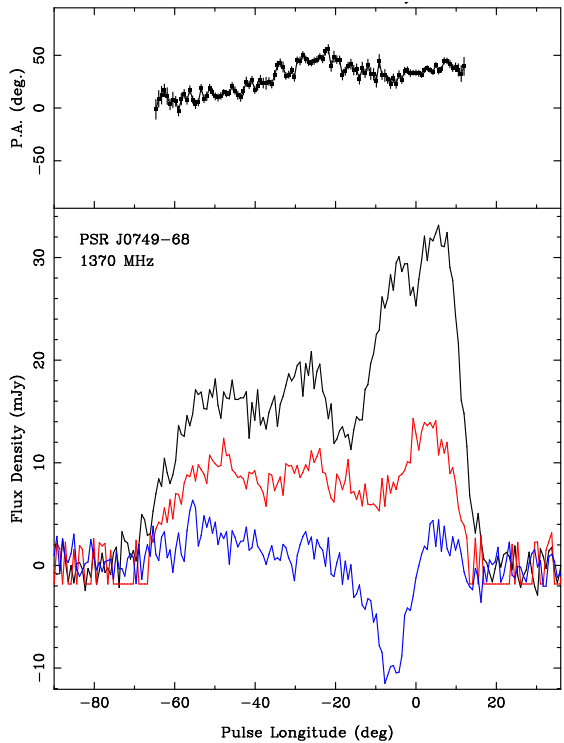}
    \vspace{-0.5cm}
    \caption{{Polarisation properties of PSR~J0749$-$68. In the upper
        panel the polarisation position angle is shown. In the lower
        panel black, blue and red denote total intensity, linear and
        circular polarisation respectively.}}\label{fig:pol}
  \end{center}
\end{figure}

\begin{figure}
  \begin{center}
     \includegraphics[scale=0.5, trim = 20mm 30mm 0mm 10mm, clip]{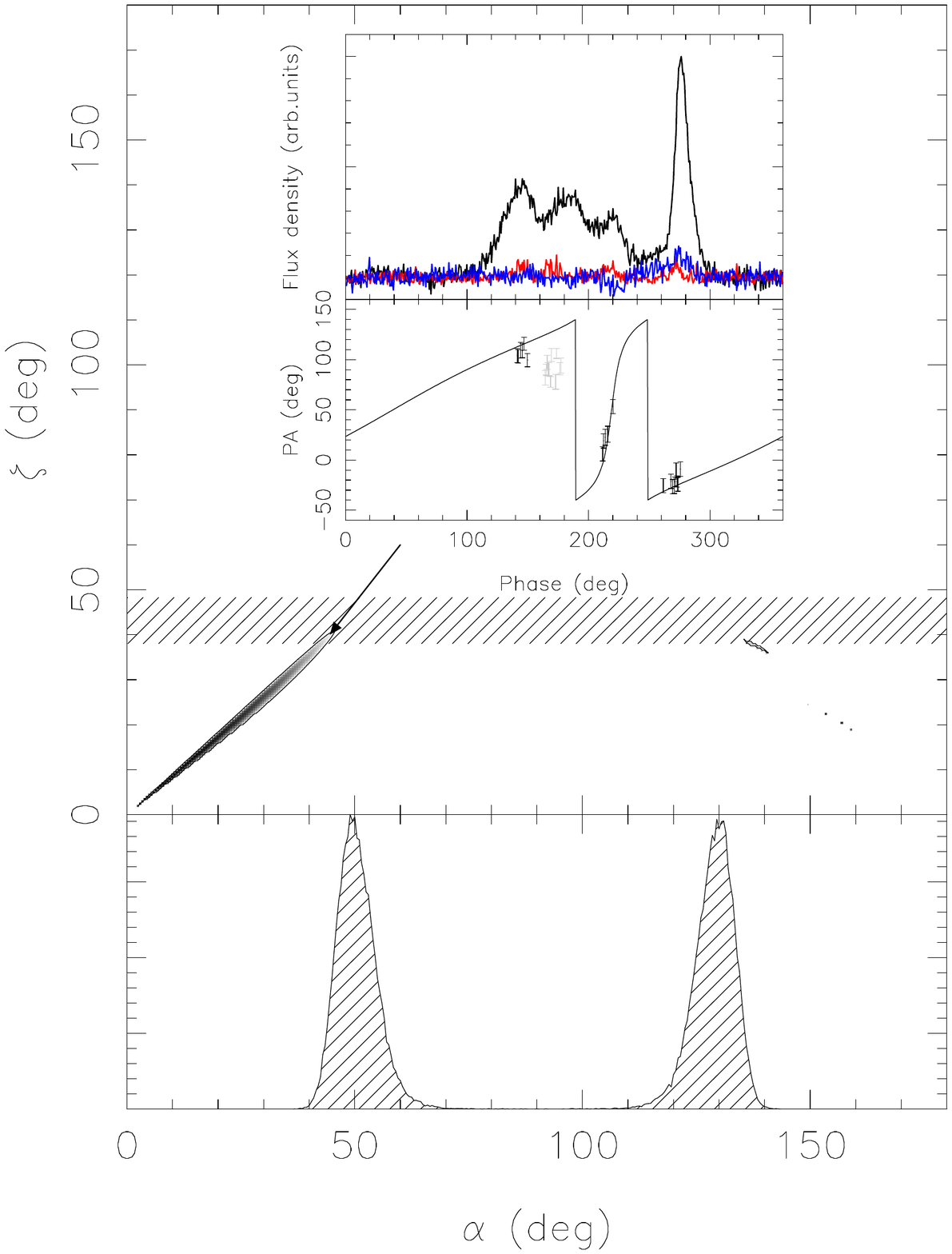}    
    \vspace{-0.5cm}
    \caption{{Polarisation properties of PSR~J1421$-$4409. The main
        top panel shows the system geometry as derived from a
        least-squares-fit of the RVM to the PA of the linearly
        polarized emission; also shown are regions of the magnetic
        inclination angle ($\alpha$) and viewing angle ($\zeta$) plane
        with the best RVM fits (1-$\sigma$ contours). We also mark the
        constraint on the orbital inclination angle as a horizontal
        strip. Also, assuming a filled emission beam, we derive a
        distribution of inclination angles (lower panel) that is
        consistent with the observed pulse width. For the point that
        satisfies the profile, polarimetric and orbital inclination
        constraints ($\alpha \, = \, 44^\circ$ and $\zeta\, = \,
        40^\circ$) we calculate the RVM PA versus phase curve and
        superimpose it on the measurements (bottom panel of
        inset). The top inset panel shows the pulse profile using the
        same conventions as in Figure~\ref{fig:pol}.}}\label{fig:pol2}
  \end{center}
\end{figure}

\textbf{PSR J0621$-$55:} This source was found in the single-pulse
pipeline and is undetectable in our periodicity searches. As such it
can be classified as a `RRAT'~\citep{km11}. As only 3 pulses have been
detected for this source it has not yet been possible to identify any
underlying periodicity.

\textbf{PSR J0749$-$68:} This pulsar is undetected in several
observations but has high S/N in others; its period averaged flux
density is $4.1$~mJy. This may be due to scintillation or possibly
nulling. The profile is broad with a width of $80\degree$. It has a
40\% linear polarization fraction which is high given the pulsar's
long spin period. Circular polarization is modest and shows a change
in sign close to the profile peak (see Figure~\ref{fig:pol}). We
measure the rotation measure to be
$-23\pm2\;\mathrm{rad}\,\mathrm{m}^{-2}$. The position angle swing is
relatively flat leading to the conclusion that the pulsar is an almost
aligned rotator. A partial timing solution has been obtained for this
pulsar so that, as indicated in Table~\ref{tab:pulsars}, its position
is determined more accurately than most of the sources presented here.

\textbf{PSR J1126$-$38:} This 887-ms pulsar, although not previously reported, appears to have also 
been independently discovered\footnote{As inferred from the survey's web pages: \texttt{http://astro.phys.wvu.edu/GBNCC/}} by the Green Bank North Celestial Cap (GBNCC) survey.

\textbf{PSR J1337$-$44:} This slow pulsar appears to be nulling and as
such is difficult to time, at least at $\sim 1.4$~GHz. Our efforts to
observe this source at $\sim 750$~MHz at Parkes (pulsars generally being
stronger at these lower frequencies, \citealt{Bates_2013}) have been scuppered due to the recent
appearance of strong terrestrial RFI in the band. The
origin of this RFI is a 4G telephone base station now located less
than $10$~km from the telescope. Given this and the Southerly declination of
the source it may evade a full timing solution for a while. The
nulling seems to be intrinsic as scintillation might be precluded as the pulsar's DM implies 
a scintillation bandwidth that is much smaller than our observing bandwidth.

\textbf{PSR J1405$-$42:} This 2.3-s pulsar is the slowest in our sample. It would 
have been missed were it not for our RFI mitigation techniques. We expect our final 
pulsar sample to contain a much higher fraction of such slow pulsars as we focus efforts 
on relevant RFI mitigation strategies as well as optimised searching, e.g. with the FFA pipeline.

\textbf{PSR J1604$-$31:} In contrast to PSR~J0749$-$68, the profile is
very narrow with a width of only $5\degr$. The fractional polarization
is low and no RM is measurable.

\textbf{PSR J1914$+$08:} This 440-ms pulsar is the one with by far the highest DM 
in our sample, as one might expect for the source that is closest to the Galactic plane. 
Although not previously reported, appears to have also 
been independently discovered\footnote{As inferred from the survey's web 
pages: \texttt{http://www.naic.edu/~palfa}} by the Pulsar Arecibo L-band Feed Array (PALFA) survey.

\textbf{PSR J2154$-$28:} This pulsar was initially missed in the F pipeline but was 
detected in the T pipeline as a result of the additional RFI mitigation performed therein.

\subsubsection{PSR J1306$-$40} 
\begin{figure*}
  \begin{center}
    \includegraphics[scale=0.6, trim = 0mm 10mm 0mm 35mm, clip]{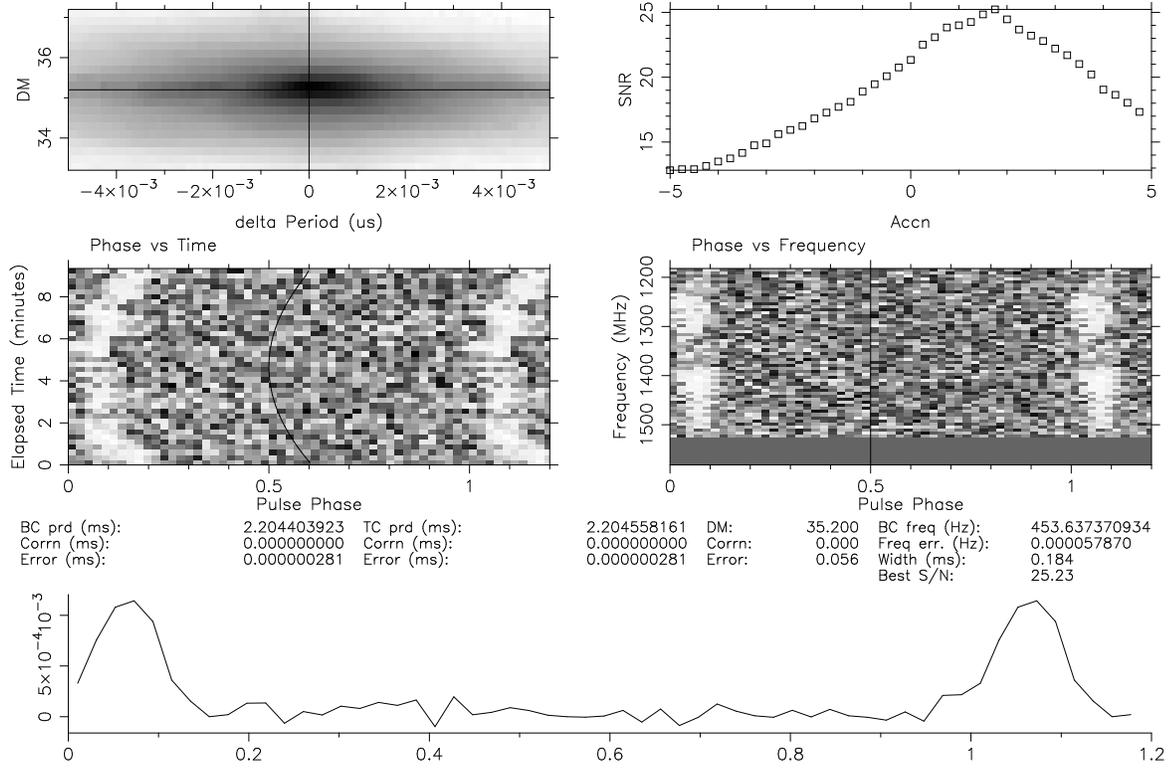}
    \caption{{One of the detections of PSR~J1306$-$40.
        The top-left panel shows the S/N (in grey scale) as a function
        of trial DM and period; the top-right panel shows S/N as a
        function of trial acceleration --- a highly significant
        orbital acceleration is evident. The middle panels the S/N
        (grey scale) as a function of time during the observation
        (left) and frequency across the observing band (right). The
        bottom panel shows the integrated pulse profile at the
        optimised period, acceleration and DM.}}\label{fig:1307}
  \end{center}
\end{figure*}

This pulsar has a period of $2.2$~ms and has proven to be the most
elusive of those reported here. It was initially detected in June 2015
in two survey pointings separated by 30 minutes (see
Figure~\ref{fig:1307}). The S/N of these initial two detections were
$\sim 14$ and $\sim 23$. The source then proved undetectable in
extended efforts to re-detect it in a total observation time of
$9$~hours. Combining the two detections we derived an improved sky
position for the source `in between' the two survey pointings, by
considering the beam model and weighting appropriately by the S/N
values. Focusing on this refined position, our best estimate for the
true position, we later re-detected the source twice again in
September 2016 in observations 5 days apart. In each detection the
signal is seen with a positive orbital acceleration where the
convention is such that this implies we are observing the pulsar on
the `near' side of its orbit. Although four detections is a small
sample this is suggestive of an ecclipsing system where the pulsar is
not detectable when on the `far' side of the orbit. The difficulty in
detecting this source is also likely compounded by scintillation. The
nominal sky position of the pulsar also happens to be within the field
where there are $130$~ks of observation accumulated, by XMM Newton, as
part of a study of a nearby Seyfert galaxy. In these data it can be
seen that the nominal position of the pulsar is coincident with the
source 3XMM~J130656.2$-$403523. The spectrum of the source is at first
glance consistent with what one might see in an ecclipsing `red back'
binary system, with a hint of variability on a $\sim 1/2$-day time
scale, but it is unclear if the spectrum is reliable given the
source's location so close to the edge of the
detector. 

\citet{Linares_2017} has further studied these X-ray data, along with
optical data from the Catalina Sky Survey, and derives a period of
$26.3$~hr. Extrapolating this period to the times at which our radio
detections have been made shows that our detections are indeed all on
the `near' side of the orbit, but with several non-detections also
falling in this range. These non-detections might be attributable to
scintillation or `transitional' behaviour in the system. Overall the
picture is consistent with the red back hypothesis. We will report in
further detail on our ongoing studies of this source in future
publications.

\subsubsection{PSR J1421$-$4409}
%
Discovered in the real-time periodicity search pipeline, PSR
J1421$-$4409 (hereon J1421) is the first millisecond pulsar discovered
by SUPERB. Its pulse profile is complex as seen in many MSPs eg. Dai
et al (2015) with half-maximum width of just $11.2\degree$ but,
because of the trailing peak, reaches the 10 percent level only after
$176\degree$ of pulse phase. The pulse averaged flux density is
$1.4$~mJy and the linear polarization is low, at just $10\%$ (see
Fig~\ref{fig:pol2}). We estimate the rotation measure to be
$-43\pm8~\mathrm{rad}\,\mathrm{m}^2$. The linear polarization loosely
tracks the total intensity profile but the circularly polarized
component becomes significant in the trailing half of the profile,
presenting a change in handedness from right to left. Individual
observations were polarization calibrated following the Measurement
Equation Template Matching technique described in
\citet{van_Straten_2013} to correct for the cross-coupling of the
feeds, using long-track observations of
PSR~J0437$-$4715. Additionally, the gain imbalance between the
receiver feeds was corrected by utilising square wave observations of
a noise diode with known polarization properties.
 
The pulsar has a 6.3-ms spin period and resides in a 30.7-day binary
system. With a minimum companion of mass of 0.18 $M_{\odot}$, J1421
would appear to be a typical PSR-HeWD binary. However, binary periods
of between 22 and 48 days are rare for MSPs \citep[the ``Camilo
  gap'';][]{Camilo_1995} with the only Galactic-field PSR-HeWD
binaries known to have orbital periods in this range being the
so-called eccentric MSPs (eMSPs, \citealt{Barr_2017}).  Although
several have been proposed, there is currently no generally accepted
model describing the evolution of eMSPs. One model of note for
discussion of J1421 is that of \citet{Antoniadis_2014}. In this model,
hydrogen shell flashes at the end of the recycling phase result in a
super-Eddington mass transfer rate between the donor and
companion. Matter then cannot be accreted onto the neutron star and
forms a circum-binary disk and it is through interaction with this
disk that eccentricity is induced in the
system. \citet{Antoniadis_2014} predicts that non-eccentric binaries
can also exist in this gap if they are capable of photo-evaporating
their circumbinary disks before they can induce eccentricity in the
orbit. A pulsar's capability to photo-evaporate its disk is
proportional to its spin-down luminosity and inversely proportional to
its semi-major axis distance. If we compare J1421's properties to
those of the eMSPs, we find that it has a slightly lower projected
semi-major axis distance than the eMSPs (12.7 lt-s as compared to a
median of 14 lt-s for the eMSPs) and its spin-down luminosity is close
to the mean of the eMSPs, $4.6\times10^{33}$ ergs s$^{-1}$. As such
J1421 does not clearly distinguish itself from the eMSPs and thus
presents a challenge to the circumbinary disk model.

\begin{table}
  \caption{Ephemerides for PSR~J1421$-$4409.}\label{tab:1421_par}
  \begin{tabular}{lr}
    \hline \hline
    Parameter & Value (Error) \\
    \hline
    Epoch (MJD)                                    & 57600                                        \\
    Pulse period, $P$ (ms)                         & 6.38572883816(3)                             \\
    Period derivative, $\dot{P}$ (10$^{-20}$)      & 1.27(4)                                      \\
    Right ascension, $\alpha$ (J2000.0)            & 14$^{\mathrm h}$21$^{\mathrm m}$20\fs9646(3) \\
    Declination, $\delta$ (J2000.0)                & $-$44\degr09\arcmin04\farcs541(4)              \\
    $\mu_\alpha$ (mas yr$^{-1}$)                   & $-$10(8)                                       \\
    $\mu_\delta$ (mas yr$^{-1}$)                   & 3(2)                                         \\
    Composite proper motion, $\mu$ (mas yr$^{-1}$) & 11(8)                                        \\
    Celestial position angle, $\phi_\mu$ (\degr)   & $-$70(2)                                       \\
    Dispersion measure, DM (pc cm$^{-3}$)          & 54.635(4)                                    \\
    Binary model                                   & ELL1                                         \\
    Solar System ephemeris                         & DE421                                        \\
    Orbital period, $P_{\rm b}$ (days)             & 30.7464535(2)                                \\
    Projected semi-major axis, $x$ (lt-s)          & 12.706655(5)                                 \\
    Orbital eccentricity, $e$                      & 0.0000128(4)                                 \\
    Epoch of periastron, $T_0$ (MJD)               & 56935.6(1)                                   \\
    Longitude of periastron, $\omega$ (\degr)      & 39(1)                                        \\
    Mass function ($\mathrm{M}_{\odot}$)                                 & 0.002330(8)                                  \\
    Characteristic age, $\tau_c$ (Gyr)             & 7.97                                        
  \end{tabular}
\end{table}

\begin{figure}
  \begin{center}
    \includegraphics[scale=0.4]{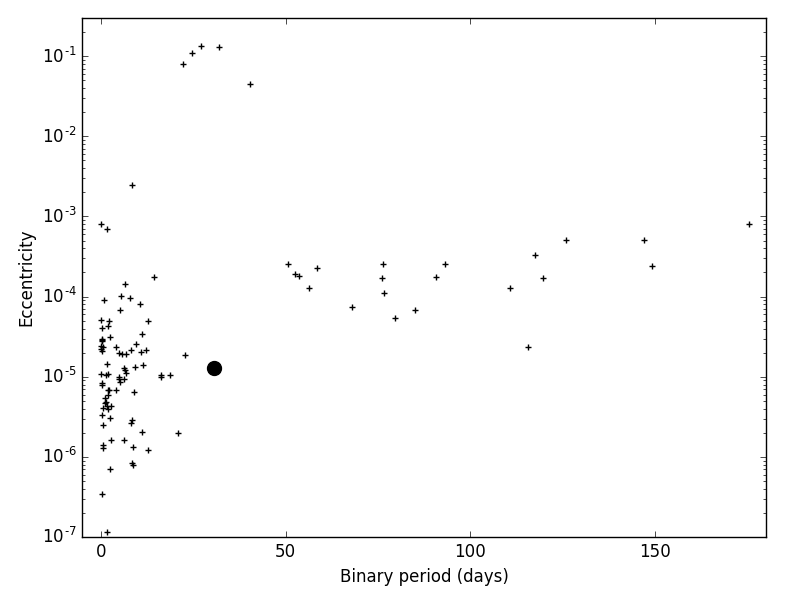}
    \caption{{Eccentricity and binary period for every binary MSP that
        is not in a double neutron star system, globular cluster and does not
        have a main sequence companion. PSR~J1421$-$4409 is identified
        with a large black circle. It can be seen that its
        eccentricity is anomalously low for systems in this period
        range, the so-called ``Camilo gap''.}}\label{fig:camilo_gap}
  \end{center}
\end{figure}

As shown in Figure~\ref{fig:camilo_gap}, J1421 falls in the previous
identified gap in eccentricity-binary period space, and is indeed more
similar to the ordinary PSR-HeWD binaries. For orbital periods larger
than 2 days, their evolution is usually well understood, allowing one
to derive an orbital period--companion mass relationship~\citep{ts99},
which appears to describe the known systems well. Assuming its
validity also for J1421, the companion mass should be $\sim
0.28M_\odot$. This relatively large value would imply a relatively low
orbital inclination angle given the value of the mass
function. Assuming a range of pulsar masses between
$1.2~\mathrm{M}_\odot$ and $1.7~\mathrm{M}_\odot$, the implied
inclination angle ranges between $38\degree$ and $48\degree$,
respectively. One can try to constrain the orbital inclination angle
also from the polarisation properties of the pulse profile. If the
pulsar's position angle (PA) swing can be described by a rotating
vector model (RVM, \citealt{rc69}), the viewing angle between the spin
axis and the line-of-sight to the observer at the closest approach to
the magnetic axis, $\zeta$, should be similar to the orbital
inclination angle $\zeta \sim i$, if the recycling process spinning up
J1421 to its current period led to the expected alignment of orbital
angular momentum and spin axis. However, fitting RVMs to recycled
pulsars is often difficult, although recent results (e.g.~Freire et
al., submitted; Berezina et al.~submitted) suggest that it is possible
in an increasing number of cases. We fitted the RVM to the PA data of
J1421 (see Figure~\ref{fig:pol2}). We use the
PSR/IEEE convention as explained in detail in \citet{vmjr10},
minimizing $\chi^2$ for a combination of $\zeta$ and the magnetic
inclination angle $\alpha$. The resulting 1-$\sigma$ contours are
shown in Figure~\ref{fig:pol2}. The best solution implies small $\alpha$
and $\zeta$ values. This is consistent with a wide pulse profile, as
observed. However, the data are consistent also with values as large
as $\sim 45\degree$. One solution is shown as a solid line in the
PA-pulse phase plot below the pulse profile shown in the inset,
representing a so-called ``inner line of sight''
(e.g.~\citealt{lk05}).  The data points in light grey have been
ignored during the fit and are shown here shifted by $90\degree$ (an
``orthogonal jump'') from their original value. Non-orthogonal jumps
are not uncommon, but given the low level of the associated linear
polarisation they were excluded. Including these data points in the
fit, does not change the best fit solution but increases the size of
the contours.

The RVM solution corresponds to a combination of angles indicated by
the arrow ($\alpha\sim44\degree$, $\zeta\sim 40\degree$). The solution
was chosen as an example for the following reasons. Assuming that the
open field-line region of the pulsar is filled with emission, the
observed pulse width can be related to $\alpha$ and $\zeta$ and the
angular radius of the open field line region, $\rho$.  There are
indications that this assumption is often not fulfilled for recycled
pulsars \citep{kxj+98}, but it can serve as a useful guide
(e.g.~Freire et al.~submitted). Assuming a period-$\rho$ scaling as
found for normal pulsars (e.g. Kramer et al. 1994), we performed
Monte-Carlo simulations that result in a distribution of $\alpha$
values consistent with the observed pulse width (see Berezina et al.,
submitted, for details). Under these assumptions, two ranges of
$\alpha$ values are consistent with the data, centred on $50\degree$
and $130\degree$, respectively, as shown in the distribution below the
$\alpha$-$\zeta$ plane. Moreover, we can also indicate the range of
inclination angles as derived from the Tauris-Savonije relationship,
assuming that $\zeta\sim i$. The corresponding range is indicated by
the horizontal hashed region. As shown by the arrow, we can find a
solution that is consistent with the data and the described
constraints. While this is not a unique solution, it does provide a
consistent picture of the evolution of the system, the pulse profile
and the polarisation information.


\subsection{Radio Frequency Interference}\label{sec:rfi}
\begin{figure*}
  \begin{center}
    \includegraphics[scale=0.25, trim = 0mm 0mm 0mm 0mm, clip]{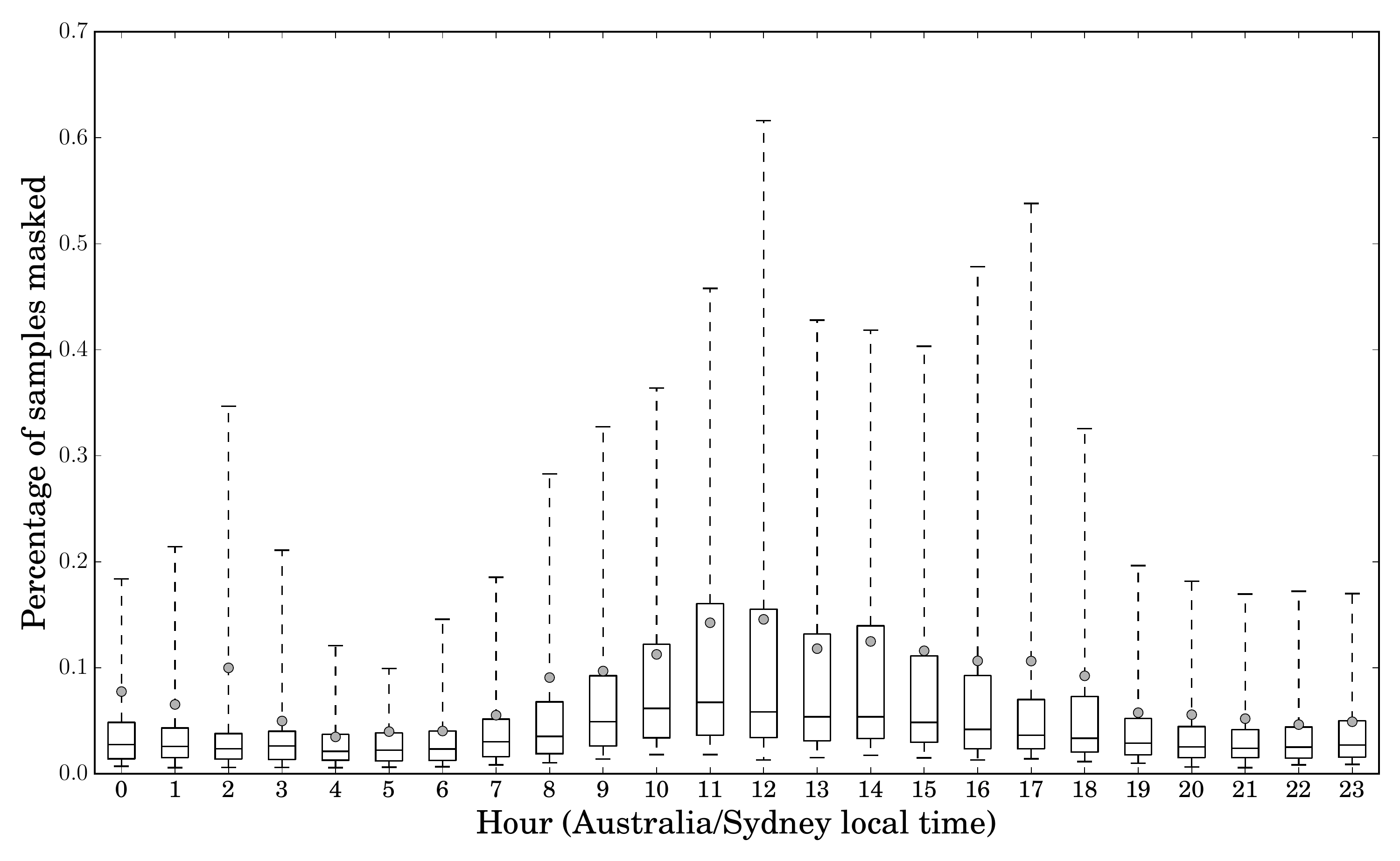}  
    \includegraphics[scale=0.25, trim = 0mm 0mm 0mm 0mm, clip]{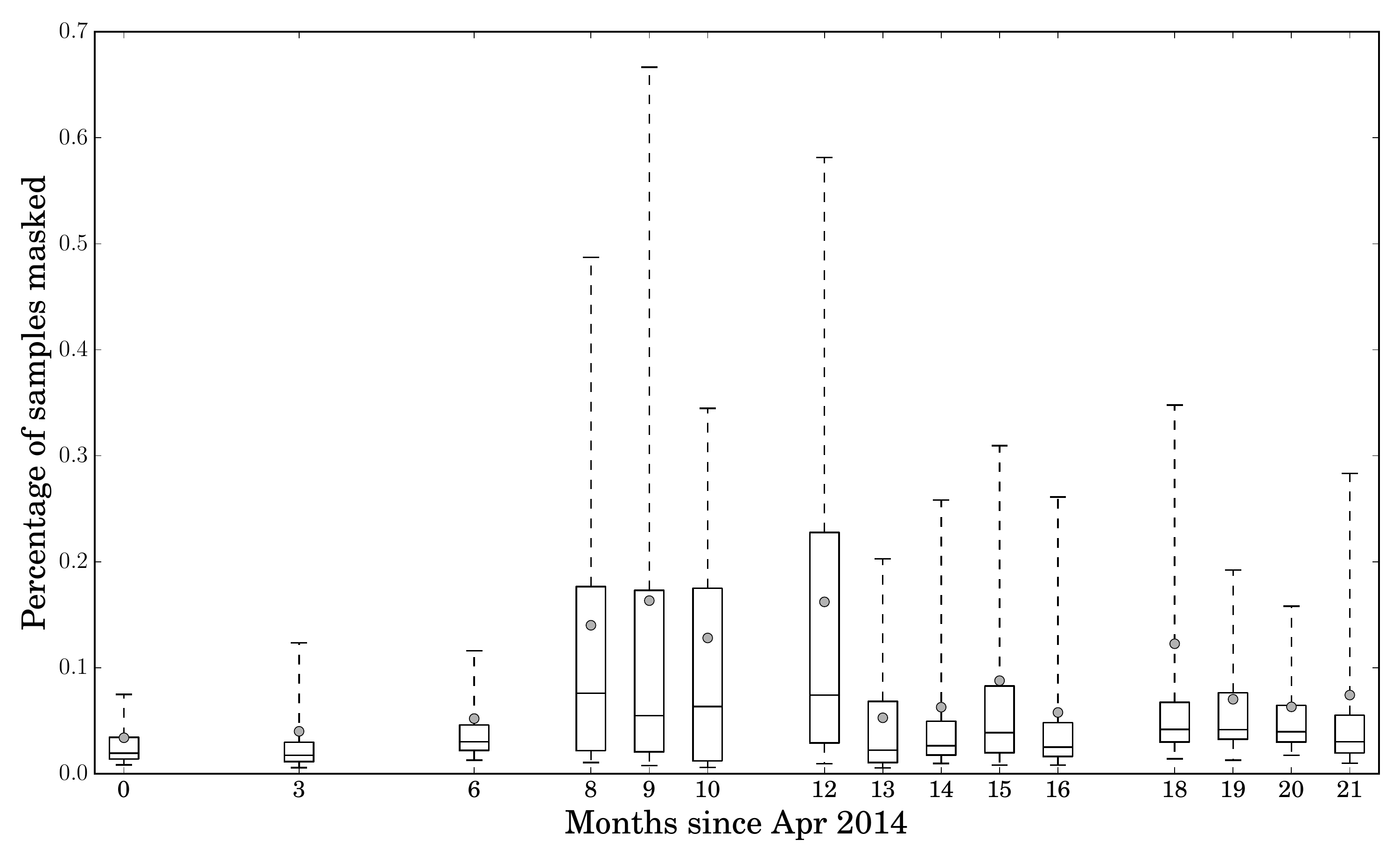}  
    \includegraphics[scale=0.25, trim = 0mm 0mm 0mm 0mm, clip]{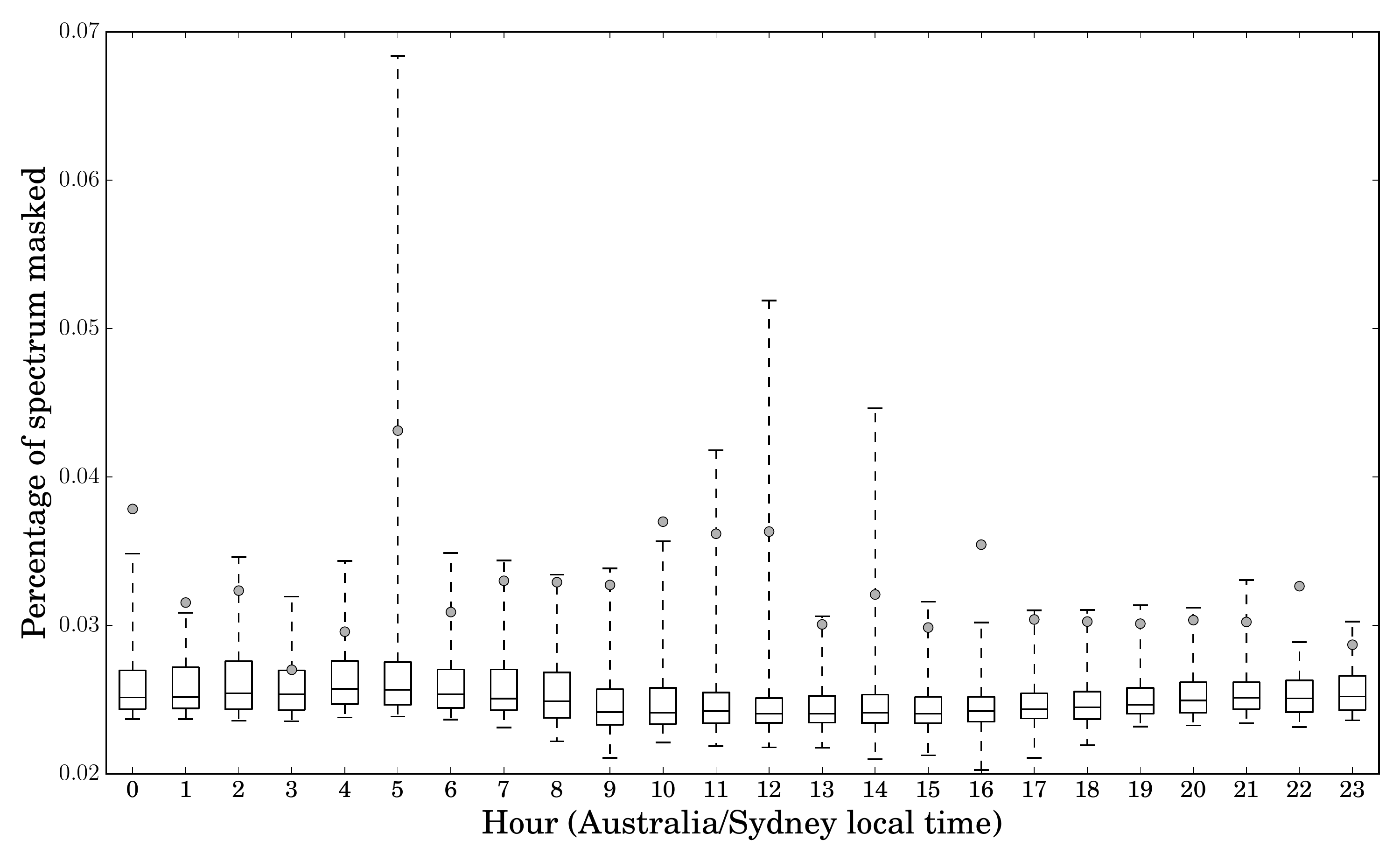}  
    \includegraphics[scale=0.25, trim = 0mm 0mm 0mm 0mm, clip]{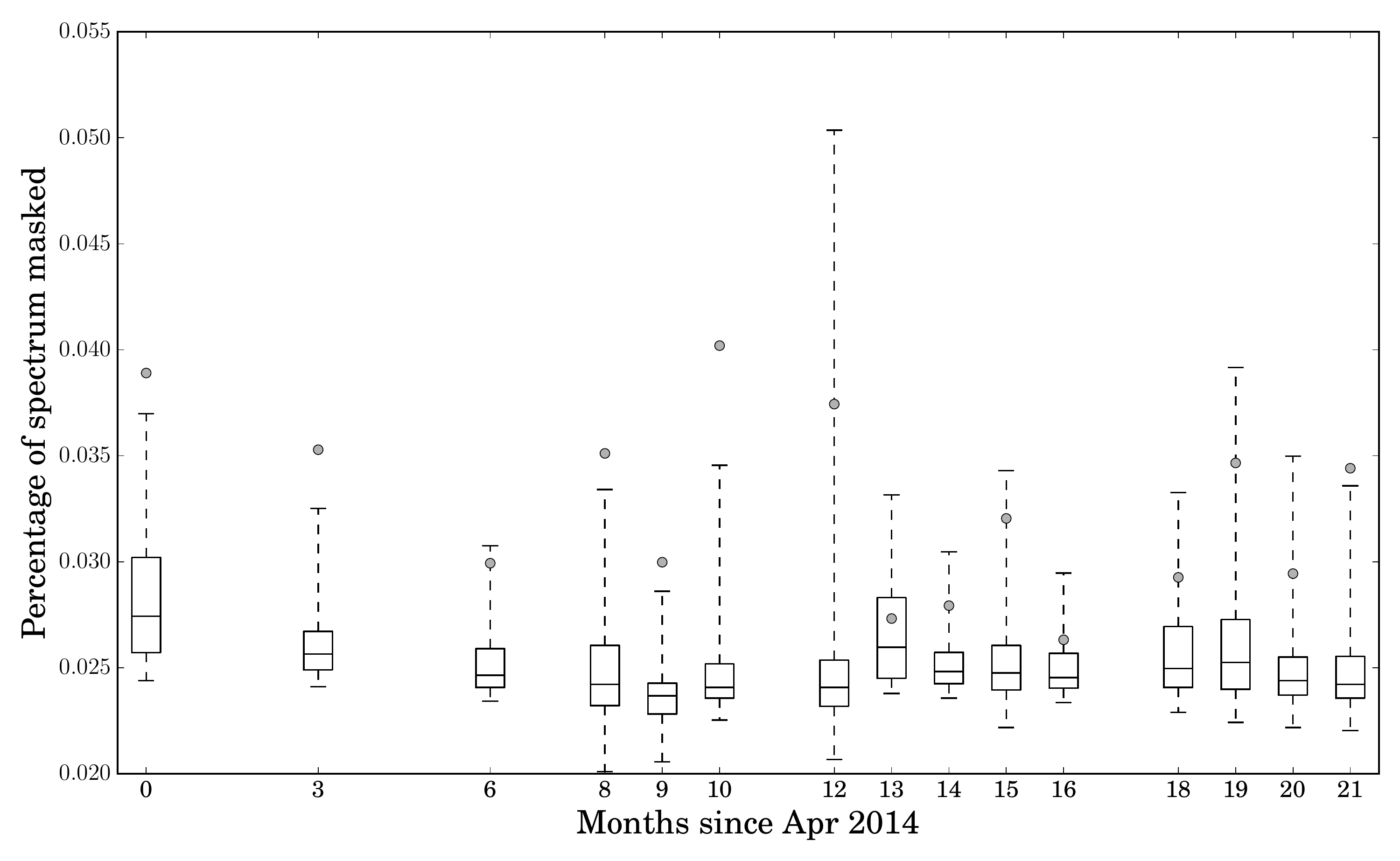}  
    \includegraphics[scale=0.25, trim = 0mm 0mm 0mm 0mm, clip]{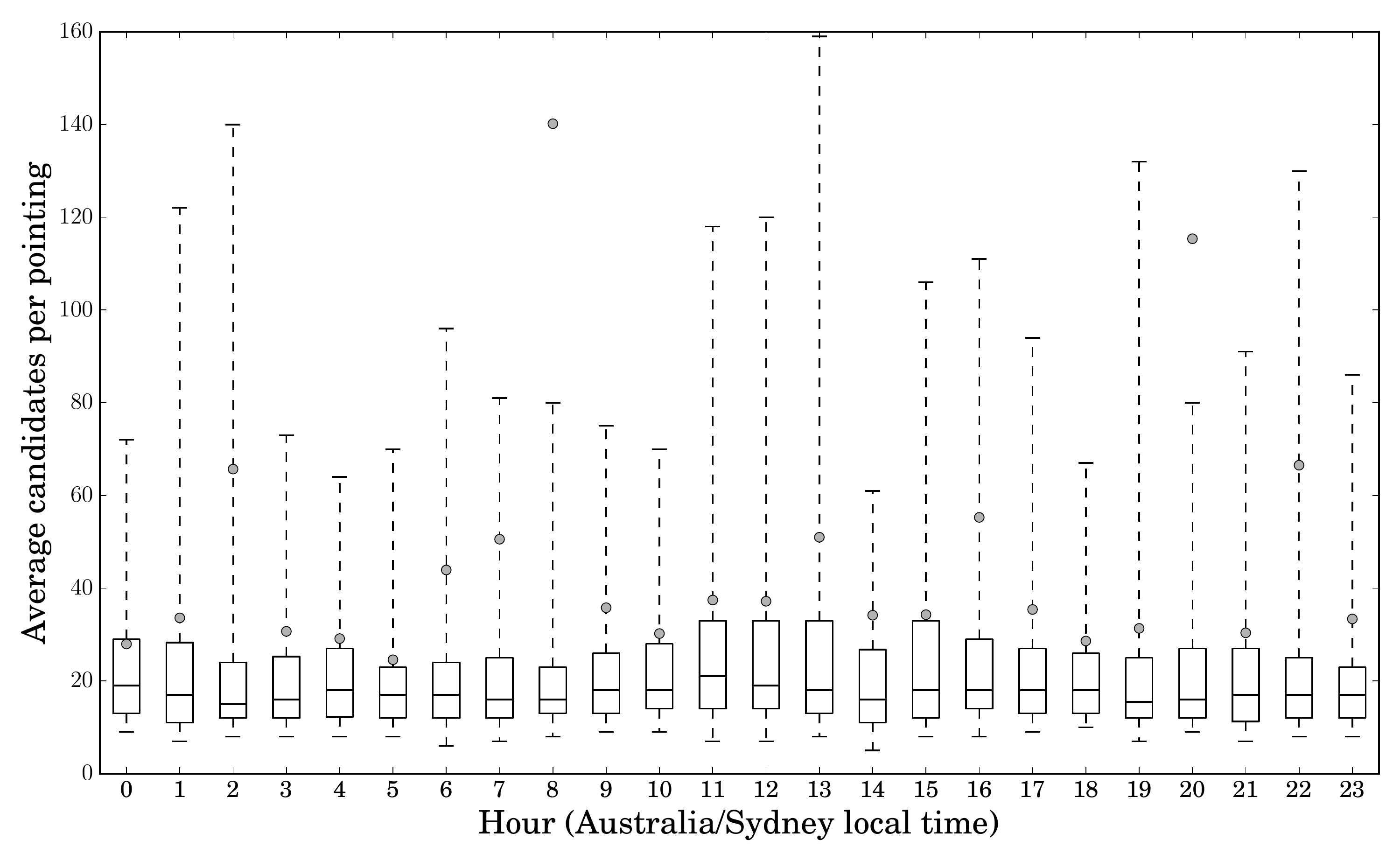}  
    \includegraphics[scale=0.25, trim = 0mm 0mm 0mm 0mm, clip]{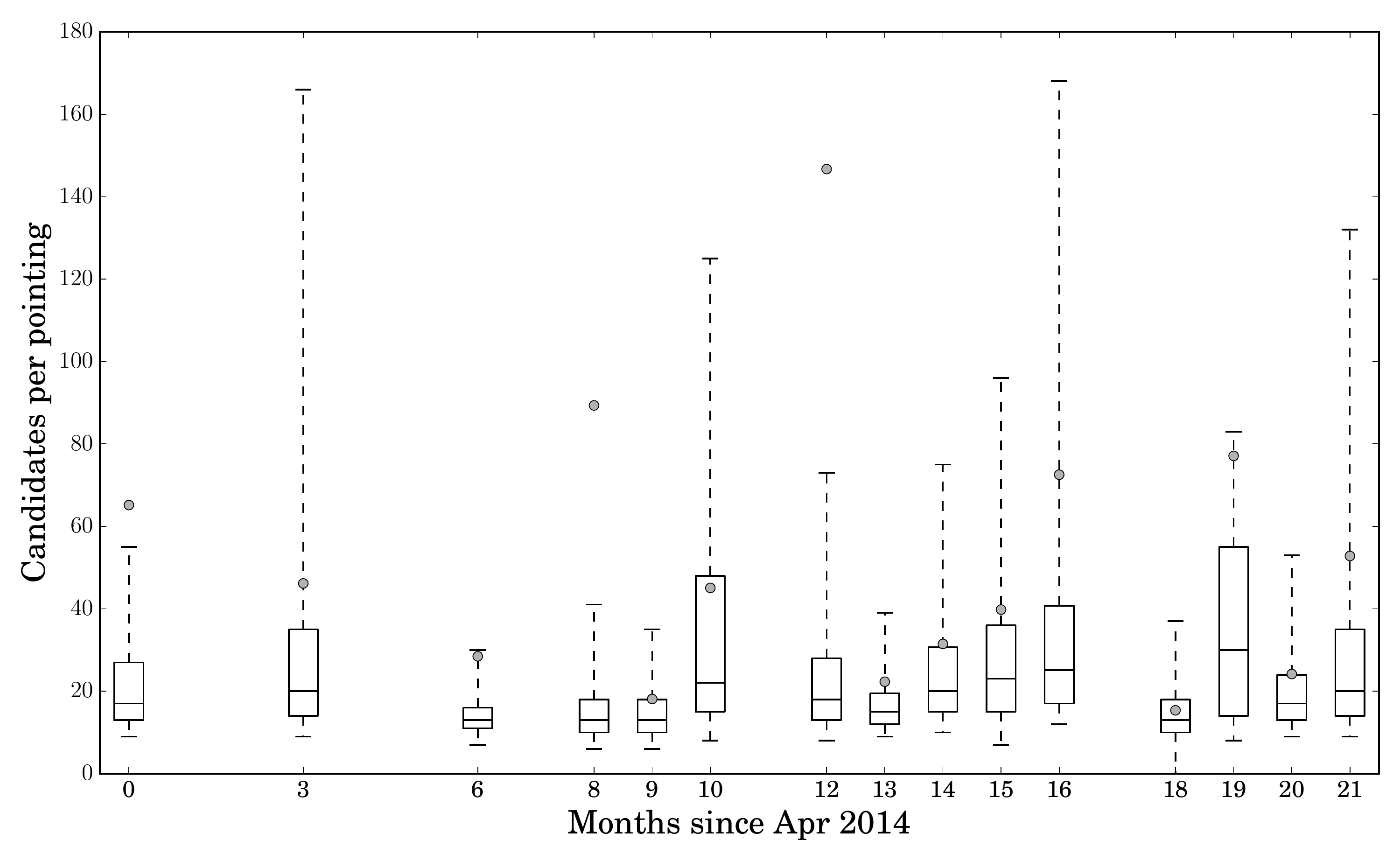}
    \caption{{Illustrations of three assessments of the RFI
        environment at Parkes, as a function of time, during the
        survey observations reported here. The metrics are shown on
        hourly (left) and monthly (right) time scales. Top row: The
        percentage of time samples masked in the eigenvector
        decomposition based assessment. Middle row: The percentage of
        spectral bins masked in the `birdie' search. Bottom row: The
        number of detected single pulse events above a threshold (more
        than six times the standard deviation in excess of the mean)
        in each pointing. These metrics are derived from a
        representative sample of 10000 random survey
        pointings.}}\label{fig:rfi}
  \end{center}
\end{figure*}

\begin{figure*}
  \begin{center}
    \includegraphics[scale=0.4, trim = 0mm 0mm 0mm 10mm, clip]{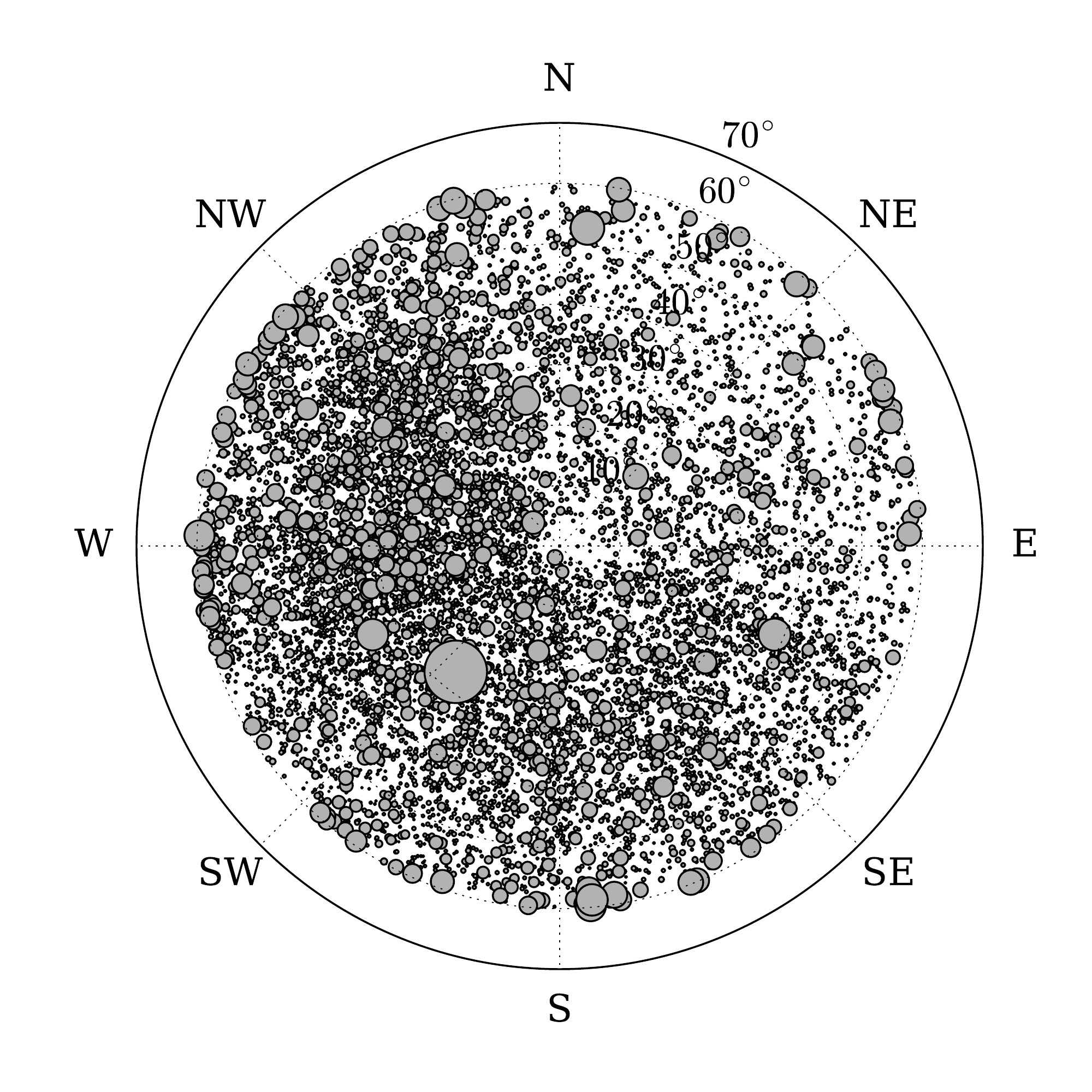}  
    \includegraphics[scale=0.4, trim = 0mm 0mm 0mm 10mm, clip]{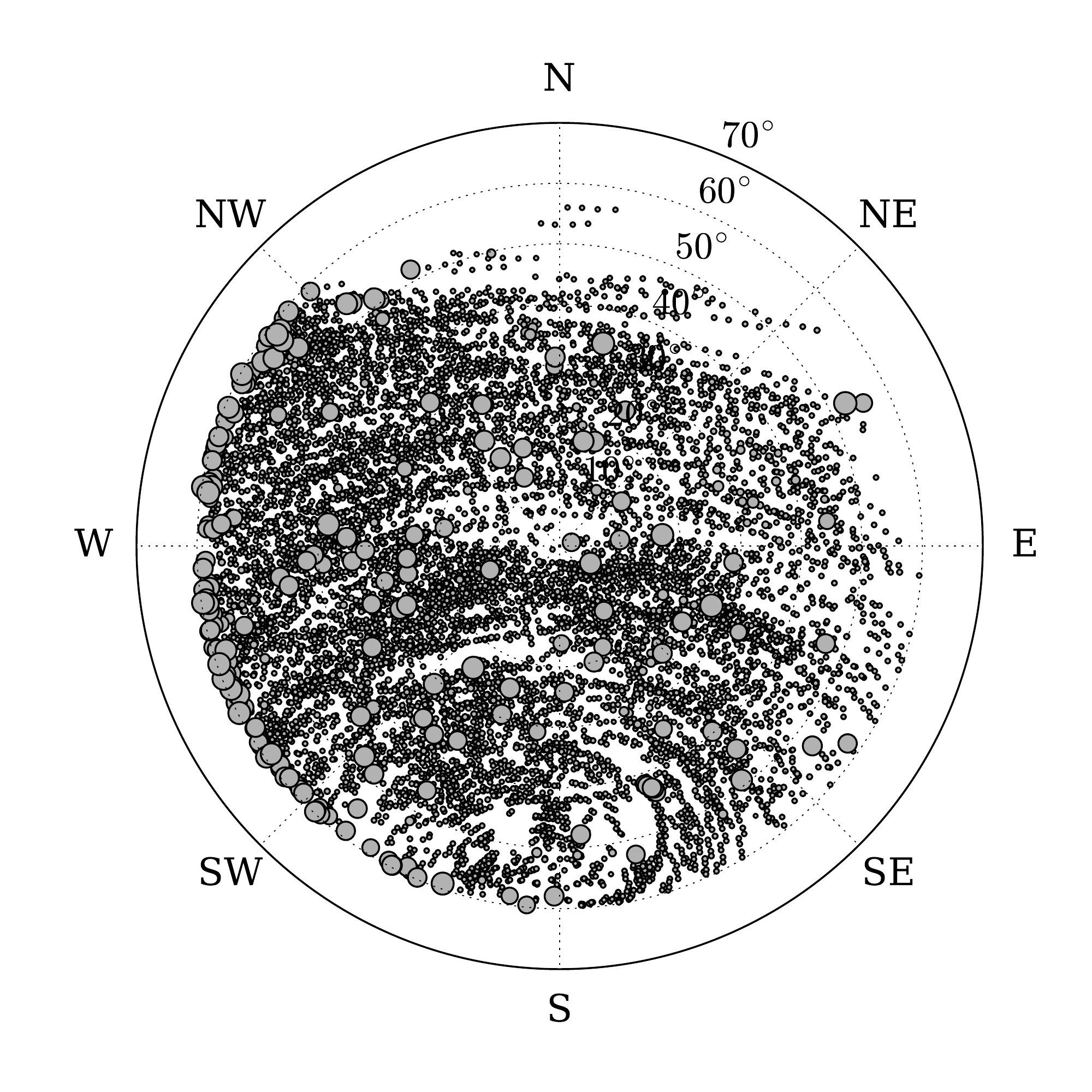}  
    \includegraphics[scale=0.4, trim = 0mm 0mm 0mm 10mm, clip]{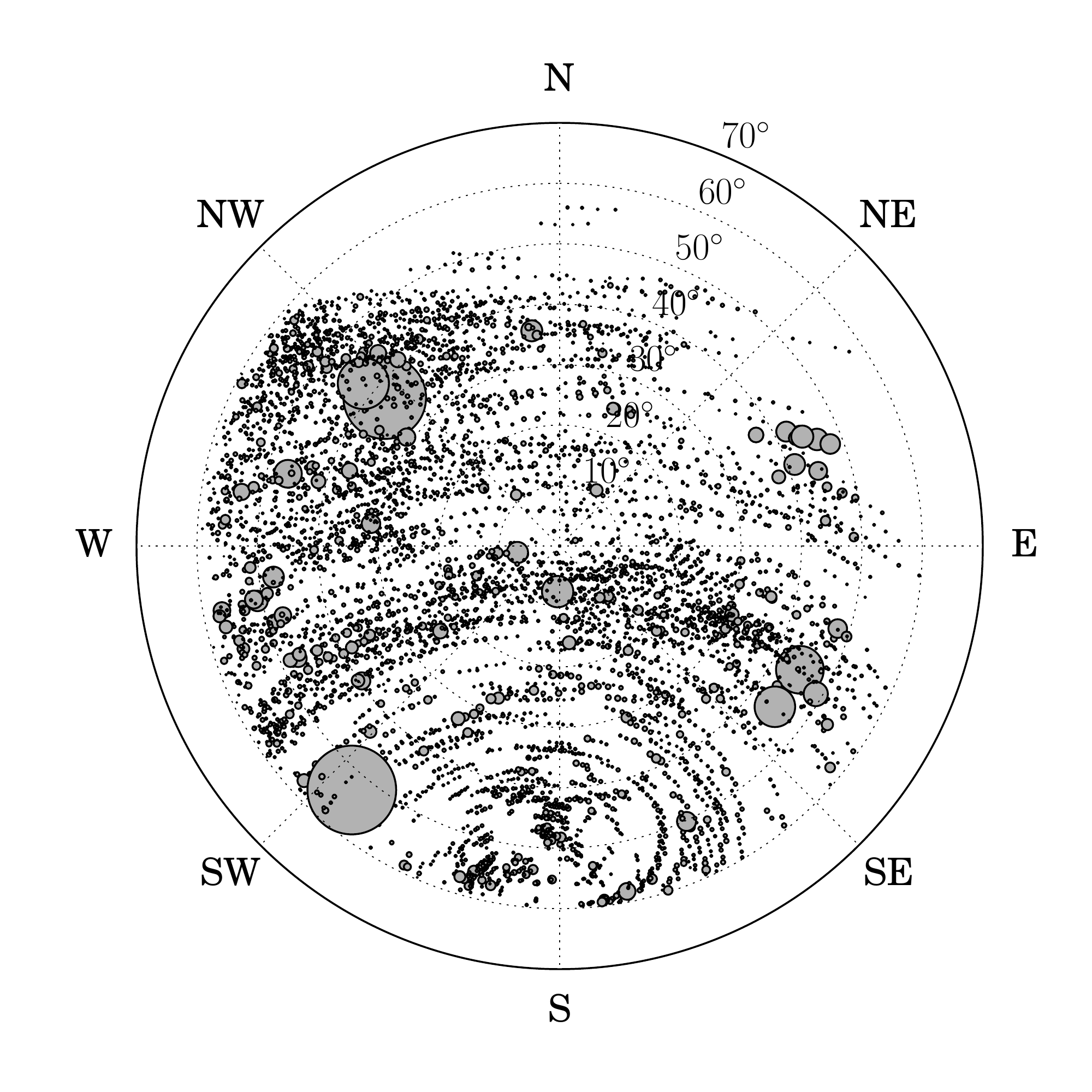}
    \caption{{Illustrations of three assessments of the RFI
        environment at Parkes, as a function of azimuth and zenith for
        the survey observations reported here. The quantities plotted
        are the same as in Figure~17: percentage masked samples in the
        top 2 panels (time samples on the left, frequencies on the
        right), and number of excess single pulse candidates in the
        bottom panel. In each panel the centre of each circle is at
        the position of the single pointing in question and the area
        of the circle denotes the magnitude of the quantity. Again,
        these metrics are derived from a random representative sample
        of 10000 survey pointings.}}\label{fig:rfi2}
  \end{center}
\end{figure*}

The survey has been subject to a large amount of RFI. This comes in
many forms --- external sources (e.g. satellites, air traffic control
radar, malfunctioning observatory equipment), internal sources
(e.g. self-induced RFI in individual beams, due to maintenance issues
with the multi-beam receiver). RFI is time variable on a number of
scales and can be both narrow and broad band. Generally speaking at
Parkes, as with most observatories worldwide, the RFI environment is
getting worse over time; even the Murchison Radio Observatory site,
perhaps the best radio astronomy site on the planet is subject to
these effects~\citep{Sokolowski_2015}. These effects have a strong
deleterious effect on our ability to detect astrophysical signals. We
are attempting to perform a census of the RFI environment at Parkes
using the SUPERB dataset. By characterising the RFI as fully as
possible we can improve the quality of our data and thus identify
otherwise obscured astrophysical signals (we have had success with
this already --- see above); this information should be useful to all
other users of the observatory also. Here, as an initial illustration
of this work, we present several metrics to quantify the effects of
RFI in our data. We examine: (i) the number of time-samples removed by
our RFI cleaning algorithms after an eigen-value decomposition of the
input from all 13 beams --- this effects both periodicity and
single-pulse searches; (ii) the number of `birdies', i.e. frequencies
in the fluctuation spectra that were removed by our algorithms ---
this effects the periodicity searches; (iii) the number of
single-pulse candidates generated --- this clearly is only a metric
relevant to single-pulse searches. The criteria used for identifying
these signals are threshold searches by comparison with the
expectation of white noise --- in the case of the birdie search an
initially `de-reddening' of the fluctuation is performed. These
metrics are illustrated in Figures~\ref{fig:rfi} and \ref{fig:rfi2}
where we examine altitude and azimuth dependence of these quantities,
and their time variability on hourly and monthly timescales. Several
patterns are evident in the data, e.g. that RFI is more prevalent
during local working hours and contamination is worst for Westerly
pointings. A thorough examination of the wide range of RFI signals in
the data will be presented in a later paper.

\section{Summary}\label{sec:last}
We have presented the features of SUPERB, an experiment designed for
searching for pulsars and fast transients using the multi-beam
receiver of the Parkes radio telescope in the 1400-MHz band. The
survey exploits a usable bandwidth of $\sim 340$ MHz, split into $390$
kHz wide channels; the integration time is 9-min and the voltages are
sampled at $2$ bits every 64 $\upmu$s. In the observations reported
here covering the time up to and including January 2016, have
accumulated an observing time-field of view product of $\sim
1350$~deg\;hr (to the half-power sensitivity level), and have
tessellated most of the sky visible from Parkes, with particular focus
on the intermediate and high Galactic latitudes.

SUPERB has introduced a few significant improvements with respect to the past large scale surveys for pulsar and/or transients carried on at Parkes or elsewhere: {\it (i)} the implementation of a real-time search for both periodic (both isolated and binary systems) and transient signals, followed by an offline deeper analysis of all the collected data; {\it (ii)} the capability of quickly distributing alerts for the occurrence of transient signals and the associated triggering of a multi-messenger campaign for follow-ups of the transient signals; {\it (iii)} the setup of a program of observations shadowing the survey pointings; {\it (iv)} the availability of full-Stokes data for transients. Observations are in agreement with the expected limiting sensitivity of the surveys. 

Here we have reported on the first 10 pulsar discoveries. One of the
two new millisecond pulsars, PSR J1421$-$4407 (6.4 ms spin period) is
in a 30.7-d orbit, showing a remarkably small eccentricity in
comparison with the other already known systems with a similar orbital
period. PSR J1306$-$40 is a 2.2-ms pulsar in an orbit that is as yet
unsolved. Indications are that it may be eclipsed by its companion (or
associated winds) for a significant fraction of its orbit, and may be
similar to the red back binary systems. Further pulsar discoveries,
with a particular focus on new ultra-long period pulsars, will be
reported in subsequent papers in this series. The next four FRB
discoveries (joining the already reported FRB~150418) will be reported
in Paper 2. The number of discoveries are roughly in agreement with
the expectations based on simple pulsar population models. 
This will be examined in detail in the future using the SUPERB dataset
to include intermittency, scintillation etc. into pulsar population
modelling for the first time. SUPERB looks set to continue working
successfully and will continue to adapt and improve over time. For
example, as new computing capabilities become available and new
telescope equipment, such as cooled phased array feeds, come into use,
the parameter space searched opens up and the survey gets ever better.

\section*{Data Access}
Data obtained in this project are archived for long-term storage on
the CASS/ANDS data server. These data are publicly available 18 months
from the day they are recorded. The data are recorded in
\textsc{sigproc} filterbank format, a \textit{de facto} standard for
pulsar search data, and are converted to \textsc{psrfits} format for
upload to the data server\footnote{\texttt{https://data.csiro.au}}. From here they can be accessed by
anybody. The full resolution data products are produced at a rate of
$\sim 46$ MiB/s when observing. For typical observing efficiencies
(considering telescope stowing due to wind, RFI or maintenance issues
that occur during routine observing) this amounts to $\sim 4$~TB per
day of observation.


\section*{Acknowledgments}
The SUPERB collaboration would like to thank S. D. Bates and
D. J. Champion for performing observations for the survey, as well as
the whole team of support staff at Parkes, and in Marsfield, for their
continuing sterling efforts essential to the success of all of the
programmes running at the facility. The Parkes radio telescope is part
of the Australia Telescope National Facility which is funded by the
Commonwealth of Australia for operation as a National Facility managed
by CSIRO. Parts of this research were conducted by the Australian
Research Council Centre of Excellence for All-sky Astrophysics
(CAASTRO), through project number CE110001020. This work was performed
on the gSTAR national facility at Swinburne University of
Technology. gSTAR is funded by Swinburne and the Australian
Government's Education Investment Fund. EP receives funding from the
European Research Council under the European Union's Seventh Framework
Programme (FP/2007-2013)/ERC Grant Agreement n. 617199. The work of MK
and RPE is supported by the ERC Synergy Grant ``BlackHoleCam: Imaging
the Event Horizon of Black Holes'' (Grant 610058).

\bibliographystyle{mnras}

\begin{thebibliography}{}

\bibitem[\protect\citeauthoryear{Ageron et~al.}{Ageron
    et~al.}{2011}]{antares} Ageron, M. et~al., 2003, Nuclear
 Instruments and Methods in Physics Research A, 656, 11

\bibitem[\protect\citeauthoryear{Antoniadis}{Antoniadis}{2014}]{Antoniadis_2014}
  Antoniadis, J., 2014, ApJ, 797, L24
  
  \bibitem[\protect\citeauthoryear{Bailes et~al.}{Bailes
    et~al.}{2011}]{Bailes_2011} Bailes M. et~al., 2011, Science, 333,
  1717

\bibitem[\protect\citeauthoryear{Bannister et~al.}{Bannister
    et~al.}{2016}]{Bannister_2016} Bannister K.~W., Stevens J.,
  Tuntsov A.~V., Walker M.~A., Johnston S., Reynolds C., Bignall H.,
  2016, Science, 351, 354

\bibitem[\protect\citeauthoryear{Barr et~al.}{Barr
    et~al.}{2017}]{Barr_2017} Barr E.~D. et~al., 2017, MNRAS, 465,
  1711

\bibitem[\protect\citeauthoryear{Barsdell, Barnes, \& Fluke}{Barsdell
    et~al.}{2010}]{Barsdell_2010} Barsdell B.~R., Barnes D.~G., Fluke
  C.~J., 2010, MNRAS, 408, 1936

\bibitem[\protect\citeauthoryear{Bates, Lorimer \& Verbiest}{Bates
    et~al.}{2013}]{Bates_2013} Bates S.~D., Lorimer D.~R., Verbiest
  J.~P.~W., 2013, MNRAS, 431, 1352




\bibitem[\protect\citeauthoryear{Bassa et~al.}{Bassa
    et~al.}{2016}]{Bassa_2016} Bassa C. G. et~al., 2016, MNRAS, 463,
  L36

\bibitem[\protect\citeauthoryear{Bernl{\"o}hr et~al.}{Bernl{\"o}hr
    et~al.}{2003}]{hess} Bernl{\"o}hr K. et~al., 2003, Astroparticle
  Physics, 20, 111

\bibitem[\protect\citeauthoryear{{Braun} et~al.}{{Braun}
    et~al.}{2015}]{Braun_2015} {Braun} R., {Bourke} T., {Green} J.~A.,
  {Keane} E., {Wagg} J., 2015, Advancing Astrophysics with the Square
  Kilometre Array (AASKA14), 174


\bibitem[\protect\citeauthoryear{{Burke-Spolaor et~al.}
    et~al.}{{Burke-Spolaor} et~al.}{2011}]{Burke_Spolaor_2011a}
  {Burke-Spolaor} S., {Bailes} M., {Ekers} R., {Macquart} J.-P.,
  {Crawford} F., III, 2011, ApJ, 727, 18

\bibitem[\protect\citeauthoryear{Burke-Spolaor et~al.}{Burke-Spolaor
    et~al.}{2011}]{BurkeSpolaor_2011} Burke-Spolaor S. et~al., 2011,
  MNRAS, 416, 2465


\bibitem[\protect\citeauthoryear{Burrows et~al.}{Burrows
    et~al.}{2005}]{swift} Burrows, D. N. et~al., 2005, Space Science
  Reviews, 120, 165

\bibitem[\protect\citeauthoryear{Caleb et~al.}{Caleb
    et~al.}{2017}]{Caleb_2017} Caleb M. et~al., 2017, MNRAS, 468, 3746

\bibitem[\protect\citeauthoryear{Cameron et~al.}{Cameron
    et~al.}{2017}]{Cameron_2017} Cameron A.~D. et~al., 2017, MNRAS,
  468, 1994
  
\bibitem[\protect\citeauthoryear{Camilo}{Camilo}{1995}]{Camilo_1995}
  Camilo F., in ``The Lives of the Neutron Stars'' (NATO ASI Series) Kluwer, 
  Dordrecht,  eds. Alpar A., Kiziloglu U., van Paradis J., p. 243.


\bibitem[\protect\citeauthoryear{Deng et~al.}{Deng
    et~al.}{2017}]{Deng_2017} Deng, X. et~al., 2017, PASA, 34, 26

\bibitem[\protect\citeauthoryear{Cordes \& Lazio}{Cordes \&
    Lazio}{2002}]{cl02} Cordes \& Lazio, 2002, astro-ph/0207156


\bibitem[\protect\citeauthoryear{Giroletti et~al.}{Giroletti
    et~al.}{2016}]{Giroletti_2016} Giroletti M. et~al., 2016, A\&A,
  593, L16

\bibitem[\protect\citeauthoryear{Hallinan et~al.}{Hallinan
    et~al.}{2007}]{Hallinan_2007} Hallinan G. et~al., 2007, ApJ, 663,
  L25

\bibitem[\protect\citeauthoryear{Horesh et~al.}{Horesh
    et~al.}{2015}]{Horesh_2015} Horesh A., Cenko S.~B., Perley D.~A.,
  Kulkarni S.~R., Hallinan G., Bellm E., 2015, ApJ, 812, 86

\bibitem[\protect\citeauthoryear{Hotan et~al.}{Hotan
    et~al.}{2004}]{hvm04} Hotan, A.~W., van Straten, W., Manchester,
  R.~N., 2004, PASA, 21, 302

\bibitem[\protect\citeauthoryear{Hyman et~al.}{Hyman
    et~al.}{2005}]{Hyman_2005} Hyman S.~D., Lazio T.~J.~W., Kassim
  N.~E., Ray P.~S., Markwardt C.~B., Yusef-Zadeh F., 2005, Nature,
  434, 50

\bibitem[\protect\citeauthoryear{Jankowski et~al.}{Jankowski
    et~al.}{2017}]{Jankowski_2017} Jankowski F. et al., 2017, MNRAS,
  in press.

\bibitem[\protect\citeauthoryear{Johnston et~al.}{Johnston
    et~al.}{2017}]{Johnston_2017} Johnston S. et al., 2017, MNRAS,
  465, 2143

\bibitem[\protect\citeauthoryear{Keane \& McLaughlin}{Keane \&
    McLaughlin}{2011}]{km11} Keane E. F., McLaughlin, M. A., 2011,
  BASI, 39, 333

\bibitem[\protect\citeauthoryear{Keane et~al.}{Keane
    et~al.}{2016}]{Keane_etal} Keane E. F. et al., 2016, Nature, 530,
  453

\bibitem[\protect\citeauthoryear{Keane}{Keane}{2016}]{keane_2016}
  Keane E. F., 2016, MNRAS, 459, 1360

\bibitem[\protect\citeauthoryear{Keith et~al.}{Keith
    et~al.}{2010}]{Keith_2010} Keith M.~J. et~al., 2010, MNRAS, 409,
  619

\bibitem[\protect\citeauthoryear{Keith et~al.}{Keith
    et~al.}{2011}]{Keith_2011} Keith M.~J. et~al., 2011, MNRAS, 419,
  1752

\bibitem[\protect\citeauthoryear{Kocz et~al.}{Kocz
    et~al.}{2010}]{Kocz_2010} Kocz J., Briggs F. H., Reynolds J.,
  2010, AJ, 140, 2086

\bibitem[\protect\citeauthoryear{Kondratiev et~al.}{Kondratiev
    et~al.}{2009}]{Kondratiev_2009} Kondratiev V.~I., McLaughlin
  M.~A., Lorimer D.~R., Burgay M., Possenti A., Turolla R., Popov
  S.~B., Zane S., 2009, ApJ, 702, 692

\bibitem[\protect\citeauthoryear{Kramer et~al.}{Kramer
    et~al.}{1994}]{kwj+94} Kramer M., Wielebinski R., Jessner A., Gil
  J. A., Seiradakis J. H., 1994, A\&AS, 107, 515
    
\bibitem[\protect\citeauthoryear{Kramer et~al.}{Kramer
    et~al.}{1998}]{kxj+98} Kramer M. et~al., 1998, ApJ, 501, 270
    
\bibitem[\protect\citeauthoryear{Kramer et~al.}{Kramer
    et~al.}{2006}]{Kramer_2006} Kramer M. et~al., 2006, Science, 312, 549

\bibitem[\protect\citeauthoryear{{Kramer} \& {Stappers}}{{Kramer} \&
    {Stappers}}{2015}]{Kramer_2015} {Kramer} M., {Stappers} B.~W.,
  2015, Advancing Astrophysics with the Square Kilometre Array
  (AASKA14), 36

\bibitem[\protect\citeauthoryear{{Leshem} \& {van der Veen}}{{Leshem}
    \& {van der Veen}}{2000}]{Leshem_2000} Leshem A., van der Veen,
  A.-J., 2000, Proceedings of ``Perspectives on Radio Astronomy:
  Technologies for Large Antenna Arrays'', ISBN: 90-805434-2-X

\bibitem[\protect\citeauthoryear{Levin et~al.}{Levin
    et~al.}{2013}]{Levin_2013} Levin L. et~al., 2013, MNRAS, 434, 1387

\bibitem[\protect\citeauthoryear{Levin et~al.}{Levin
    et~al.}{2010}]{Levin_2010} Levin L. et~al., 2010, ApJ, 721, L33

\bibitem[\protect\citeauthoryear{Linares}{Linares}{2017}]{Linares_2017}
  Linares, M., 2017, MNRAS submitted, astro-ph/1707.00698

\bibitem[\protect\citeauthoryear{Lorimer \& Kramer}{Lorimer
    \& Kramer}{2005}]{lk05} Lorimer D.~R., Kramer M., 2005, Handbook of Pulsar Astronomy, CUP

\bibitem[\protect\citeauthoryear{Lorimer et~al.}{Lorimer
    et~al.}{2007}]{Lorimer_2007} Lorimer D.~R., Bailes M., McLaughlin
  M.~A., Narkevic D.~J., Crawford F., 2007, Science, 318, 777

\bibitem[\protect\citeauthoryear{Lyne et~al.}{Lyne
    et~al.}{1990}]{Lyne_1990} Lyne A.~G. et~al., 1990, Nature, 347,
  650

\bibitem[\protect\citeauthoryear{Manchester et~al.}{Manchester
    et~al.}{1990}]{mhth05} Manchester, R.~N., Hobbs, G.~B., Teoh, A.,
  Hobbs, M., 2005, AJ, 129, 1993

\bibitem[\protect\citeauthoryear{McLaughlin et~al.}{McLaughlin
    et~al.}{2006}]{McLaughlin_2006} McLaughlin M.~A. et~al., 2006,
  Nature, 439, 817

\bibitem[\protect\citeauthoryear{Morello et~al.}{Morello
    et~al.}{2014}]{Morello_2014} Morello, V., Barr, E. D., Bailes, M.,
  Flynn, C. M., Keane, E. F., van Straten, W., 2014, MNRAS, 443, 1651

\bibitem[\protect\citeauthoryear{{Morello}}{{Morello}}{2016}]{vincent_thesis}
  {Morello} V., 2016, MSc Thesis, Swinburne University of Technology,
  {``Discovering Pulsars with Machine Learning''}



\bibitem[\protect\citeauthoryear{Osten \& Bastian}{Osten \&
    Bastian}{2008}]{Osten_2008} Osten R.~A., Bastian T.~S., 2008,
  {ApJ}, 674, 1078

\bibitem[\protect\citeauthoryear{{Petroff} et~al.}{{Petroff}
    et~al.}{2014}]{Petroff_2014} {Petroff} E. et~al., 2014, ApJ, 789,
  L26

\bibitem[\protect\citeauthoryear{{Petroff} et~al.}{{Petroff}
    et~al.}{2015}]{Petroff_2015a} {Petroff} E. et~al., 2015, MNRAS,
  447, 246

\bibitem[\protect\citeauthoryear{Petroff et~al.}{Petroff
    et~al.}{2015}]{Petroff_2015b} Petroff E. et~al., 2015, MNRAS, 451,
  3933

\bibitem[\protect\citeauthoryear{{Petroff} et~al.}{{Petroff}
    et~al.}{2016}]{FRBCAT} {Petroff} E. et~al., 2016, PASA,
  33, 45

\bibitem[\protect\citeauthoryear{{Price} et~al.}{{Price}
    et~al.}{2017}]{Price_2016} {Price} D.~C. et~al., 2016,
  J. Astron. Instrum., 05, 1641007

\bibitem[\protect\citeauthoryear{{Radakrishnan} \& {Cooke}}{{Radakrishnan}
    \& {Cooke}}{1969}]{rc69} Radakrishnan V., Cooke D. J., 1969, Astrophys. Lett., 3 225 

\bibitem[\protect\citeauthoryear{{Ransom} et~al.}{{Ransom}
    et~al.}{2002}]{Ransom_2002} {Ransom} S.~M. et~al., 2002, AJ, 124,
  1788

\bibitem[\protect\citeauthoryear{Ransom}{Ransom}{2005}]{Ransom_2005}
Ransom S.~M., 2005, Science, 307, 892

\bibitem[\protect\citeauthoryear{{Raza} \& {van der Veen}}{{Raza} \&
    {van der Veen}}{2002}]{Raza_2002} Raza J. \& van der Veen A.~J.,
  2002, IEEE Signal Processing Letters, 9, 64

\bibitem[\protect\citeauthoryear{Rickett}{Rickett}{1970}]{Rickett_1970}
Rickett B.~J., 1970, MNRAS, 150, 67

\bibitem[\protect\citeauthoryear{Robinson et~al.}{Robinson
    et~al.}{1995}]{Robinson_1995} Robinson C., Lyne A.~G., Manchester
  R.~N., Bailes M., D{\textquotesingle}Amico N., Johnston S., 1995,
  MNRAS, 274, 547

\bibitem[\protect\citeauthoryear{Roy et~al.}{Roy
    et~al.}{2010}]{Roy_2010} Roy J., Gupta Y., Pen Ue-Li., Peterson
  J. B., Kudale S., Kodilkar J., 2010, ExA, 28, 25

\bibitem[\protect\citeauthoryear{Sokolowski et~al.}{Sokolowski
    et~al.}{2015}]{Sokolowski_2015} Sokolowski M., Wayth R. B.,
  Morgan, L., 2015, proceedings of GEMCCON, 2015 IEEE Global;
  doi:/10.1109/GEMCCON.2015.7386856

\bibitem[\protect\citeauthoryear{Spitler et~al.}{Spitler
    et~al.}{2016}]{Spitler_2016} Spitler L.~G. et~al., 2016, Nature,
  531, 202
  
\bibitem[\protect\citeauthoryear{van Straten et~al.}{van Straten et~al.}{2010}]{vmjr10} 
van Straten W., Manchester R.~N. Johnston S., Reynolds J.~E., 2010, PASA,
  27, 104

\bibitem[\protect\citeauthoryear{van Straten}{van Straten}{2013}]{van_Straten_2013} van Straten W., 2013, ApJS,
  204, 13

\bibitem[\protect\citeauthoryear{{Staelin}}{{Staelin}}{1969}]{Staelin_1969}
  {Staelin} D.~H., 1969, IEEE Proceedings, 57, 724

\bibitem[\protect\citeauthoryear{Tauris \& Savonije}{Tauris \&
    Savonije}{1999}]{ts99} Tauris T.~M., Savonije G.~J., 1999, A\&A,
  350, 928

\bibitem[\protect\citeauthoryear{Thornton et~al.}{Thornton
    et~al.}{2013}]{Thornton_2013} Thornton D. et~al., 2013, Science,
  341, 53

\bibitem[\protect\citeauthoryear{Tiburzi et~al.}{Tiburzi
    et~al.}{2013}]{Tiburzi_2013} Tiburzi C. et~al., 2013, MNRAS, 436,
  3557

\bibitem[\protect\citeauthoryear{{Tingay} et~al.}{{Tingay}
    et~al.}{2013}]{mwa} Tingay, S. J. et~al., 2013, PASA, 30, 7


\bibitem[\protect\citeauthoryear{Williams \& Berger}{Williams \&
    Berger}{2016}]{WilliamsBerger} Williams P. K. G., Berger E., 2016,
  ApJ, 821, L22

\bibitem[\protect\citeauthoryear{Yao et~al.}{Yao et~al.}{2017}]{ymw16}
  Yao, M. J., Manchester, R. N., Wang, N., 2017, ApJ, 835, 29


\end{thebibliography}

\end{document}